\documentclass[useAMS,usenatbib]{mn2e}

\usepackage[dvips]{graphicx}  
\usepackage{aas_macros}  
\usepackage{amsmath}
\usepackage{chngcntr}
\usepackage[usenames]{color}

\newcommand{\Msolar}{\mbox{\,$\rm M_{\odot}$}}        
\newcommand{\Rsolar}{\mbox{\,$\rm R_{\odot}$}}        
\newcommand{\Lsolar}{\mbox{\,$\rm L_{\odot}$}}        

  \newcommand{\ion}[2]{\mbox{\,#1\,{\sc #2}}}         
  \newcommand{\kmsec}{\,\mbox{$\mbox{km}\,\mbox{s}^{-1}$}}    
 \newcommand{\cmss}{\,\mbox{$\mbox{cm}\,\mbox{s}^{-2}$}}    
  \def\simge{\mathrel{\raise1.16pt\hbox{$>$}\kern-7.0pt
    \lower3.06pt\hbox{{$\scriptstyle \sim$}}}}           
  \def\simle{\mathrel{\raise1.16pt\hbox{$<$}\kern-7.0pt
    \lower3.06pt\hbox{{$\scriptstyle \sim$}}}}           
\title[V652~Her: radial velocities]{Subaru and {\it Swift} observations of V652~Herculis: 
resolving the photospheric pulsation\thanks{Based in part on data collected at Subaru Telescope, which is operated by the National Astronomical Observatory of Japan} }
\author[C.S. Jeffery et al.]{C.S. Jeffery$^{1,2}$, D. Kurtz$^3$,  H. Shibahashi$^4$, R.L.C. Starling$^5$, V. Elkin$^3$,
\newauthor  P. Monta\~n\'es-Rodr\'{\i}guez$^{6,7}$, and J. McCormac$^8$  \\
$^{1}$Armagh Observatory, College Hill, Armagh BT61 9DG, UK\\
$^{2}$School of Physics, Trinity College Dublin, College Green, Dublin 2, Ireland\\
$^{3}$Jeremiah Horrocks Institute, University of Central Lancashire, Preston PR1 2HE, UK\\
$^{4}$Department of Astronomy, School of Science, The University of Tokyo, Bunkyo-ku, Tokyo 113-0033, Japan \\
$^{5}$Department of Physics and Astronomy, University of Leicester, University Road, Leicester LE1 7RH, UK\\
$^{6}$Instituto de Astrof\'{i}sica de Canarias (IAC), V\'{i}a L\'{a}ctea s/n E-38200, La Laguna, Spain\\
$^{7}$Departamento de Astrof\'{i}sica, Universidad de La Laguna, E-38206 La Laguna, Spain\\
$^{8}$Department of Physics, University of Warwick, Gibbet Hill Road, Coventry, CV4 7AL, UK\\
}
\begin{document}

\date{Accepted \ldots. Received \ldots; in original form \ldots}

\pagerange{\pageref{firstpage}--\pageref{lastpage}} \pubyear{2014}

\maketitle

\label{firstpage}

\begin{abstract}
High resolution spectroscopy with the Subaru High Dispersion Spectrograph, and {\it Swift}
ultraviolet photometry are presented for the pulsating extreme helium star V652\,Her. {\it Swift}
provides the best relative ultraviolet photometry obtained to date, but shows no direct evidence for
a shock at ultraviolet or X-ray wavelengths. Subaru has provided high spectral and high temporal
resolution spectroscopy over 6 pulsation cycles (and eight radius minima). These data have enabled a
line-by-line analysis of the entire pulsation cycle and provided a description of the pulsating
photosphere as a function of optical depth. They show that the photosphere is compressed radially by
a factor of at least two at minimum radius, that the phase of radius minimum is a function of
optical depth and the pulse speed through the photosphere is between 141 and 239 \kmsec\ (depending
how measured) and at least ten times the local sound speed. The strong acceleration at minimum
radius is demonstrated in individual line profiles; those formed deepest in the photosphere show a
jump discontinuity of over 70\kmsec\ on a timescale of 150\,s. The pulse speed and line profile
jumps imply a shock is present at minimum radius. These empirical results provide input for
hydrodynamical modelling of the pulsation and hydrodynamical plus radiative transfer modelling of
the dynamical spectra.
\end{abstract}


\begin{figure}
	\centering
		\includegraphics[width=0.47\textwidth]{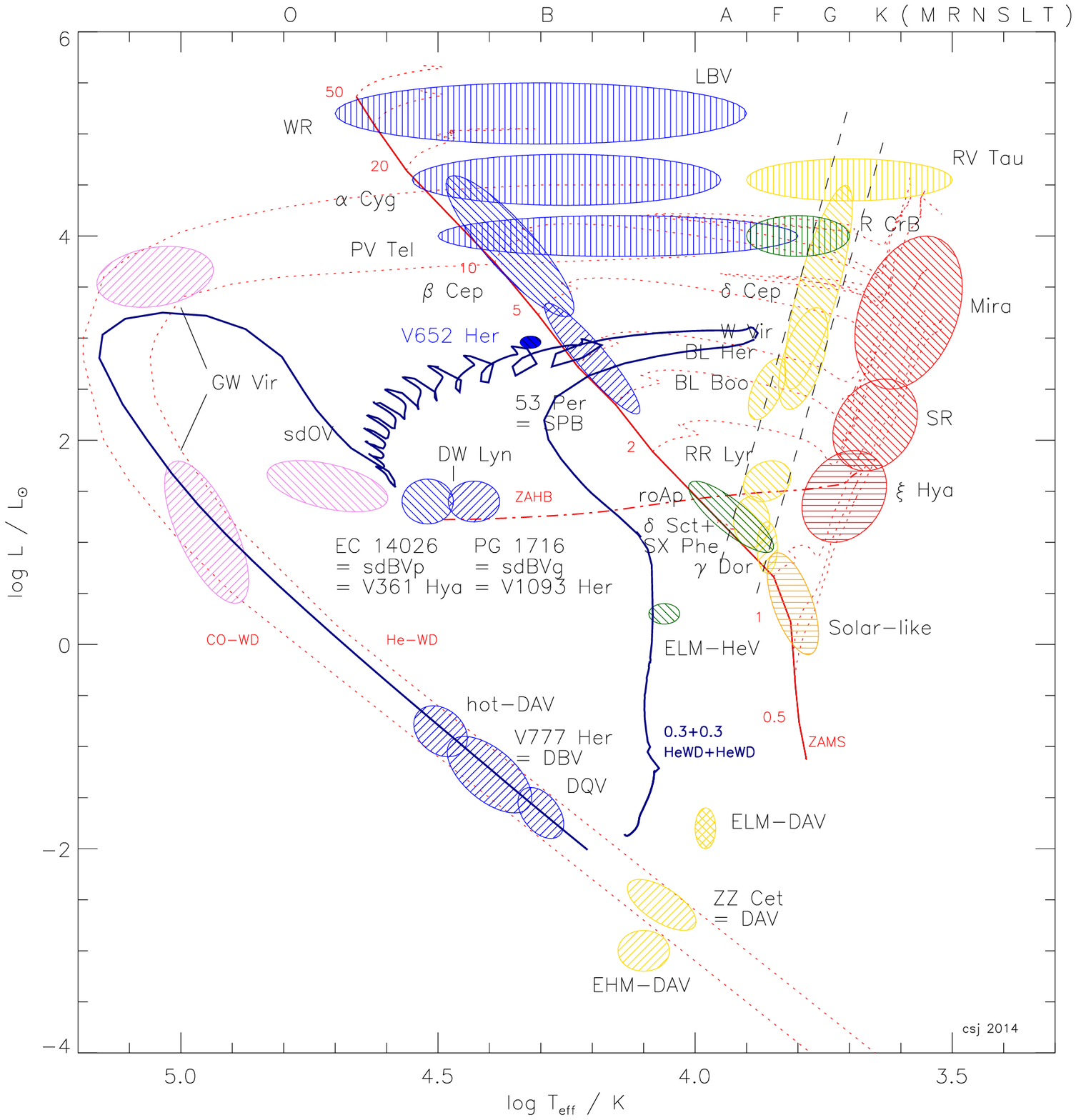}
\caption{Luminosity-effective temperature (or Hertzsprung-Russell) diagram showing the position of
V652\,Her \citep{jeffery01b} and the post-merger evolution track for a double helium white dwarf
binary (0.3+0.3 \Msolar: thick blue line: \citealt{zhang12a}). The latter shows expansion to become
a yellow giant following helium-shell ignition. As the helium-burning shell migrates towards the
core, the star contracts through a series of loops caused by helium-shell flashes, until it reaches
the helium main-sequence. On completion of core helium burning, the star returns to the white dwarf
cooling track. Although V652\,Her is located below the main-sequence in the HR diagram, it is
spectroscopically a giant; historically, the latter is a spectroscopic term and depends on
luminosity-to-mass ratio (or surface gravity). In the case of the 0.6 \Msolar V652\,Her, this is
higher (lower) than for main-sequence stars of the same effective temperature. Also shown are the
approximate locations of other pulsating variables coloured roughly by spectral type, the zero-age
main sequence and horizontal branch, the Cepheid instability strip, and evolution tracks for model
stars of various masses, indicated by small numbers (\Msolar). Shadings represent opacity-driven
p-modes ($\backslash\backslash\backslash$), g-modes (///) and strange modes ($|||$) and
acoustically-driven modes ($\equiv$). Approximate spectral types are indicated on the top axis.
Based on figures by J. Christensen-Dalsgaard and subsequently by \citet{jeffery08.coast}. }
	\label{f:hrd}
\end{figure}

\begin{table}
\caption{Published surface, pulsation and other properties of V652\,Her. 
Users should refer to original publications. 
Notes. 1:\citet{kilkenny05}, 2:\citet{jeffery01b}, 3:\citet{przybilla05}, 4:\citet{jeffery99}.  }
\label{t:pars}
\begin{tabular}{l rr ll}
\hline
$P_0$ (1974) & \multicolumn{2}{c}{$0.107\,992\,82\pm0.000\,000\,02$}   & d & 1 \\
$\dot{P}/P$ & \multicolumn{2}{c}{$-7.426\pm0.005\times10^{-8}$} & d\,d$^{-1}$ & 1 \\[2mm]
$\langle T_{\rm eff} \rangle$ (ion equil) & 22\,930 & $\pm 10$ & K & 2 \\
$\langle \log g \rangle$ (He {\sc i} lines) & 3.46 & $\pm 0.05$ & $\cmss$ & 2 \\
$\langle T_{\rm eff} \rangle$ (total flux) & 20\,950 & $\pm 70$ & K & 2 \\
$\langle R \rangle$  & 2.31 & $\pm 0.02$ & \Rsolar & 2 \\
$\langle L \rangle$ & 919 & $\pm 014$ & \Lsolar & 2 \\
$M$ & 0.59 & $\pm 0.18$ & \Msolar & 2 \\
$d$ & 1.70 & $\pm 0.02$ & kpc & 2 \\[2mm]
$T_{\rm eff}$ (NLTE) & 22\,000 & $\pm 500$ & K & 3 \\
$\log g $ (NLTE) & 3.20 & $\pm 0.10$ & $\cmss$ & 3 \\
$n_{\rm H}$ (NLTE ) & 0.005 & $\pm 0.0005$ &   & 3 \\[2mm]
$T_{\rm eff}$ (mean) & 24\,550 & $\pm 500$ & K & 4 \\
$\log g $ (mean) & 3.68 & $\pm 0.10$ & $\cmss$ & 4 \\
$\log n_{\rm H}$  & --2.16 & $\pm 0.07$ &  & 4 \\
$\log n_{\rm He}$  & 0 &      &  & 4 \\
$\log n_{\rm C}$  & --4.40 & $\pm 0.27$ &  & 4 \\
$\log n_{\rm N}$  & --2.61 & $\pm 0.06$ &  & 4 \\
$\log n_{\rm O}$  & --4.00 & $\pm 0.08$ &  & 4 \\
$\log n_{\rm Fe}$  & --4.14 & $\pm 0.15$ &  & 4 \\
\hline
\end{tabular}
\end{table}

\section{Introduction}
\label{s:intro}
Just over fifty years ago, \citet{berger63.v652} used the Palomar 5-m (200-inch) telescope to show
that the early B star BD$+13^{\circ}3224$ is hydrogen deficient and a probable subdwarf. Whilst all
permitted neutral helium lines were visible, the hydrogen Balmer lines were unexpectedly weak,
although still detectable. A decade later, \citet{landolt75} discovered light variations with a
period 0.108\,d and an amplitude of 0.1~mag (in $V$) in a star lying outside any known pulsation
instability strip. Spectroscopic confirmation that BD$+13^{\circ}3224$ (V652\,Herculis;
\citealt{IBVS77.1248}) is a radially pulsating variable caused as much interest as the shape of its
radial velocity curve \citep{hill81,lynasgray84,jeffery86}. The latter shows a very rapid
acceleration around minimum radius followed by a constant deceleration covering nine tenths of the
pulsation cycle. The maximum surface acceleration reaches some $300 {\rm \,m\,s^{-2}}$, or $30g$ in
terrestrial terms, and poses several questions that this paper addresses. To set these in context,
we first reprise our current knowledge of V652\,Her.

Additional multiwavelength photometry and ultraviolet observations from the {\it International
Ultraviolet Explorer (IUE)} enabled the radius and distance to be obtained \citep{hill81,lynasgray84} using
the Baade-Wesselink method \citep{baade26,wesselink46}. The best radius measurement to date is
$2.31\pm0.02 ~ {\rm R_{\odot}}$ which, with a surface gravity measured from the spectrum, gives a
mass of $0.59\pm0.18\,{\rm M_{\odot}}$ \citep{jeffery01b}. Analysis of the spectrum shows a surface
that is mostly normal in metal abundances, except for carbon and oxygen, which are weak, and
nitrogen, which is sufficiently overabundant to demonstrate that the surface helium is the product
of nearly complete CNO burning \citep{jeffery99}. The 0.5~per~cent contamination by hydrogen
represents a potential puzzle \citep{przybilla05}.

An apparent error in the predicted time of minimum radius \citep{hill81}, prompted
\citet{kilkenny82} to show that the pulsation period had been decreasing since its first
measurement, with subsequent observations of the light curve demonstrating up to third-order
behaviour in the ephemeris \citep{kilkenny84,kilkenny88,kilkenny91,kilkenny05}. The linear term is
consistent with a secular radial contraction of some 30\,km\,yr$^{-1}$, assuming a standard period
-- mean density relation.
 
At the time of discovery, no known mechanism could excite the pulsations in V652\,Her
\citep{saio86}. Not even the strange modes which drive pulsations in more luminous B-type helium
stars would work for V652\,Her \citep{saio88b}. However, the development of improved opacities for
the Sun and other stars and the discovery of additional opacity from iron group elements at $\approx
2\times10^5$K \citep{opal92,OP94} provided an explanation; pulsation driving is provided through the
classical $\kappa$-mechanism \citep{saio93}. Non-linear calculations subsequently reproduced the
period, light and velocity curves with satisfactory precision \citep{fadeyev96,montanes02}.
 
In seeking an evolutionary origin for V652\,Her, single star models are unable to explain all of the
observed properties, although \citet{jeffery84b} argued that a star contracting towards the helium
main-sequence could match the mass, luminosity and contraction rate. { Such configurations 
might arise as a result of a late or very-late helium-core flash 
in a star that has already left the giant branch \citep{brown01}. Such a star could 
share  the observed properties of V652\,Her (Table\,\ref{t:pars}). 
For the surface to be extremely hydrogen-deficient, a binary companion
would be necessary to remove most of the red-giant envelope  either by common-envelope ejection
 or by stable mass transfer. 
In the case of V652\,Her, there is no evidence for a stellar-mass companion \citep{kilkenny96}.
It is possible that high-order terms in the pulsation ephemeris might be caused by light-travel  
delays across an orbit around a planetary-mass companion having a period $8 - 12$\,y \citep{kilkenny96,kilkenny05}. 

The most successful model for the origin of V652\,Her is }
that of the merger of two helium white dwarfs \citep{saio00}; following such a merger, the ignition
of helium in a shell at the surface of the accretor and at the base of the accreted material creates
a yellow giant, which subsequently contracts at a rate and luminosity commensurate with observation
(Fig.~\ref{f:hrd}), ultimately to become a helium-rich hot subdwarf \citep{zhang12a}. 
 
There might, therefore, seem to be little left to learn from V652\,Her (Table\,\ref{t:pars}). 
However, in obtaining the
highest precision measurement of the radial velocity curve to date, \citet{jeffery01b} noted that
the extreme acceleration at minimum radius could be associated with a shock travelling outwards
through the photosphere. { \citet{jeffery01b} identified possible line splitting in 
strong lines around minimum radius as supporting evidence for the idea of a shock. 
\citet{fadeyev96} predicted that a shock at minimum radius should be associated 
with a spike in the light curve; none has been detected in visible light \citep{kilkenny05} 
and the ultraviolet  light curve is too  sparsely sampled to show such a feature 
\citep{lynasgray84,jeffery01b}. }
Since the absence of hydrogen renders the stellar photosphere much less
opaque than is normally the case, sufficiently high precision observations would offer unique
opportunities (a) to  establish whether a shock does indeed occur and (b) to explore the effect of
pulsation on the photosphere itself. Since previous applications of the Baade-Wesselink method have
relied on model atmospheres in hydrostatic equilibrium, higher precision data would also provide the
opportunity to implement high accuracy dynamical models of the envelope and photosphere.

Another question prompted by the almost uniform surface acceleration between radius minima is how
close the surface motion is to a ballistic trajectory. A simple picture of V652\,Her likens the
pulsating surface to a rocket that is rapidly accelerated upwards, after which it is in free fall
until it returns to its original state near minimum radius, at which point the cycle repeats. The
physics of the atmosphere of such a ``rocket star'' is of considerable interest.

Scrutiny of the published parameters for V652\,Her (Table\,\ref{t:pars}) raises further questions.
Application of the classical period--mean-density relation $P \sqrt{\bar{\rho}/\bar {\rho_{\odot}}} = Q$ \citep{shapley14} 
using the \citet{jeffery01b}
mass and radius and with a pulsation constant $Q\approx 0.033$\,d 
 \citep{fadeyev96} indicates an unacceptably long period of 0.15 d and implies 
 that the measured surface gravity is too low by a factor of 2. Appealing to other measurements of
the surface gravity, \citet{przybilla05} found a substantially lower value (at maximum radius) even though their
model techniques were more sophisticated, while analyses using older methods  had
found higher values \citep{jeffery99}. Reconciling the spectroscopic and pulsation parameters for 
V652\,Her represents a challenge for both theory and observation. 

In this paper we present a unique set of spectroscopic observations of V652\,Her. We describe the
spectrum and present the radial velocity and equivalent width measurements obtained for each line.
Using these data we explore the radial velocity behaviour of the photosphere in terms of its
vertical structure. We explore the behaviour of line profiles throughout the pulsation cycle, and
especially around minimum radius, including both strong and weak lines. We also present observations
aimed at detecting evidence for a shock at ultraviolet and X-ray wavelengths. The result is a unique
empirical description of the behaviour of the outer layers of a pulsating star, resolved over nearly
three decades of optical depth.

\begin{figure}
	\centering
		\includegraphics[height=0.47\textwidth,angle=90]{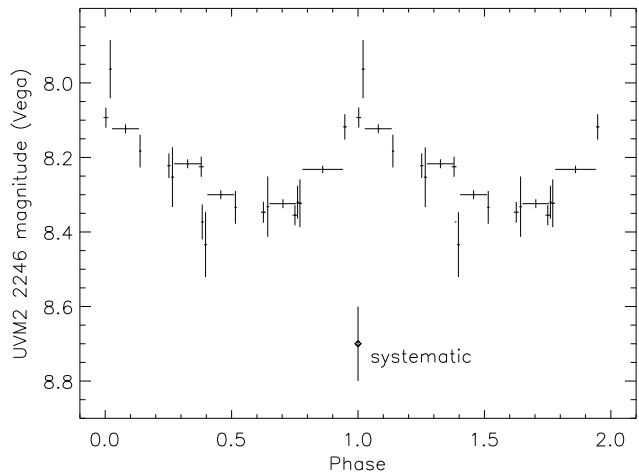}
\caption{{\it Swift} UVOT photometry of V652\,Her from 2010 May 6, phase folded using the ephemeris given in \S\,4.2. 
The systematic uncertainty is marked separately. } 
	\label{f:uvot}
\end{figure}

\begin{figure}
	\centering
		\includegraphics[height=0.47\textwidth,angle=90]{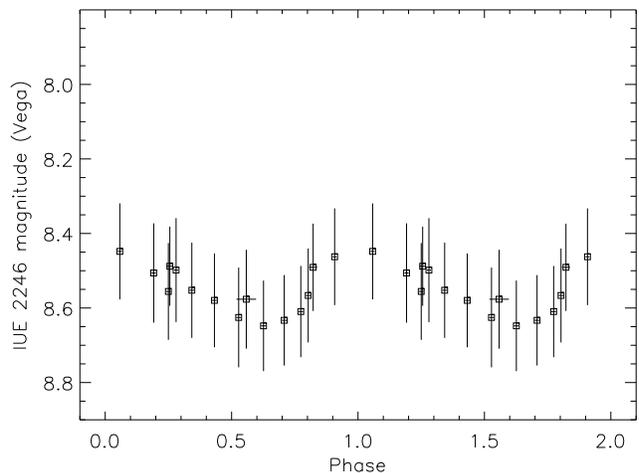}
\caption{{\it IUE} photometry of V652\,Her re-extracted at $\lambda_{\rm c}=2246$~\AA, $\Delta \lambda =498$~\AA, and phase folded using the ephemeris given in \S\,4.2. } 
	\label{f:iue}
\end{figure}

\begin{figure}
	\centering
		\includegraphics[height=0.47\textwidth,angle=90]{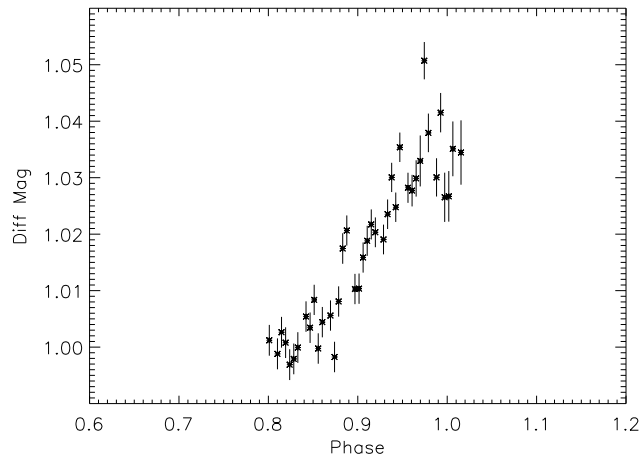}
\caption{NGTS visual photometry of V652\,Her from 2010 May 6, phased using the ephemeris given in \S\,4.2.} 
	\label{f:ngts}
\end{figure}

\section[]{Observations: {\it Swift} and NGTS photometry}

Observations were made with all instruments on board the gamma-ray burst detecting satellite {\it
Swift} \citep{gehrels04} on 2010 May 6 during a target of opportunity programme (obsID 00031714).
Six observations, with a total exposure of $\sim$7~ks, were carried out at intervals intended to
cover different phases of the light curve.  { V652\,Her was not detected at high energies, 
as expected, with a 3$\sigma$ upper
limit on the 0.3-10 keV X-ray count rate of $1.25 \times 10^{-3}$ count s$^{-1}$ (using the Bayesian
method) for all six observations combined. } It was only detected with the UV/Optical telescope
(UVOT) observing in the UVM2 filter ($\lambda_{\rm c}=2246$~\AA). V652\,Her is close to the bright
limit for the UVOT and therefore a special calibration was required to extract reliable photometry
\citep{page13}. This calibration has been applied; the resulting photometry in Vega magnitudes and
phased to the most recent ephemeris (\S\,\ref{s_ephem}) is shown in Fig.~\ref{f:uvot}. A systematic
uncertainty of $\pm0.1$ mag should be added to these measurements, as deemed necessary by
\citet{page13}, who suggest that the method be limited to stars fainter than ninth magnitude, while
V652~Her is brighter than this by a few tenths or so. However, the photometry is considered accurate
to within the $1\sigma$ errors shown.

There is a discrepancy between the {\it Swift} UVOT photometry obtained in 2010 and the {\it IUE}
photometry obtained between 1979 and 1984 as reported by \citet{lynasgray84} and \citet{jeffery01b}.
The photometry has been re-extracted to form equivalent Vega magnitudes in a bandpass centred around
2246~\AA\ having the same FWHM (full width at half maximum) as the UVM2 filter (498~\AA); the
results are shown in Fig.\,\ref{f:iue}. It will appear from these data that, modulo the systematic
error in the UVOT zero point, V652\,Her may have brightened at 2246~\AA\ by $\approx$0.5~mag in the
30~yr since first observed at these wavelengths.

Recall from \S\,\ref{s:intro} that the observed change in pulsation period is consistent with a
secular radial contraction of some $30 {\rm \,km\,yr^{-1}}$ for a star with a radius of 2.3\Rsolar\
and a mean effective temperature (in 2000) around 23\,000\,K. Assuming a constant luminosity (for
the sake of argument, but consistent with evolution models), such a contraction would correspond to
an increase in effective temperature of some $100 {\rm \,K\,yr^{-1}}$. Considering the 30~yr elapsed
between the {\it IUE} and UVOT observations, the effective temperature of V652\,Her should have increased
by some 3000\,K. Under the constant luminosity assumption, this implies a dimming by
$\approx$0.15~mag at 2246~\AA, and brightening at shorter wavelengths. It is therefore not, at
present, clear where the UVOT/{\it IUE} discrepancy arises.

{ Broad-band optical photometry was obtained on the same day as the {\it Swift} observations, with the
Next Generation Transit Survey (NGTS) prototype\footnote{www.ngtransits.org/prototype.shtml} on La Palma 
\citep{mccormac14},  a 20\,cm Takahashi telescope with a 1k$\times$1k e2v CCD
operating in the bandpass $600-900$\,nm. The NGTS-P data were bias subtracted and flat-field
corrected using {\sc PyRAF}\footnote{{\sc PyRAF} is a product of the Space Telescope Science
Institute, which is operated by AURA for NASA.} and the standard routines in {\sc IRAF}\footnote{{\sc IRAF} 
is distributed by the National Optical Astronomy Observatories, which are operated by the Association of Universities for Research in Astronomy, Inc., under cooperative
agreement with the National Science Foundation.} and aperture photometry was performed using {\sc DAOPHOT} \citep{stetson87}.  }
Only relative fluxes were obtained, since the bandpass is custom made and standard stars were not
observed. Due to weather conditions, only the part of the light curve leading to and including
visual light maximum was observed. Phased to the same ephemeris, these data are shown in
Fig.~\ref{f:ngts} and, as far as is possible for these data, confirm the ephemeris is correct for
the date of the {\it Swift} observations.

\section[]{Observations: Subaru spectroscopy}

Observations were obtained with the High Dispersion Spectrograph (HDS) \citep{noguchi02} of the Subaru
telescope on the nights of 2011 June 6 and 7 (Julian dates 2455719--20). A total of 772 spectra were
obtained in pairs to cover the wavelength ranges $398.7-482.2$~nm and $487.3-570.6$~nm (386 spectra
in each range), all with an exposure time of 120\,s and spectral resolution of $R = 90\,000$. The
mean time between exposures for read out of the detectors was $\approx$56~s. The median
signal-to-noise (S/N) ratio in the continuum region around $452-453$~nm was 73 on the first night
and 58 on the second night. At the epoch of these observations, the exposure time corresponds to
0.013 pulsation cycles, whilst the mean sampling interval corresponds to $\approx$0.020 cycles,
which had slightly unfortunate consequences for the phase distribution.

Thorium-argon (ThAr) comparison lamp exposures were obtained at the beginning and middle of each
night. For wavelength calibration, the ThAr spectrum obtained in the middle of each night was used.
The CCD images were processed using {\small IRAF} and {\small ESO-MIDAS} software to extract and
merge the \'echelle orders to obtain one-dimensional (1D) spectra.

The overscan data were processed using {\small IRAF} and a {\tt cl} script from the Subaru website
(http://subarutelescope.org/\-Observing/\-Instruments/\-HDS/\-index.html). This script also
subtracted the average bias from each frame. To extract and merge the \'echelle orders to 1D, the
fits files were converted to {\small ESO-MIDAS} internal format. Each 2D frame was rotated so that
the \'echelle orders were approximately horizontal with wavelength increasing along each order from
left to right and order number increasing from top to bottom (as viewed on screen). For each
observing night an average image was constructed using 50 object frames obtained in the middle of
the night. These average images were used to determine the order positions. Order detection was done
using a Hough transform \citep{ballester94}. The {\small ESO-MIDAS} command for order definition
also provides an estimate of the background in the interorder space. \'Echelle orders were extracted
by taking a sum of pixel values over the slit with one pixel width running along the orders. The
length of the slit was defined using the average object frame.

Each extracted order was normalized to a smooth continuum. This smooth continuum was obtained using
a polynomial approximation to the median-filtered extracted orders, avoiding spectral lines. When an
order has strong lines with broad wings, the continuum for that order was calculated by
interpolation between neighbouring orders.

For wavelength calibration, ThAr spectra were extracted from 2D images. Spectral lines were detected
in the extracted orders and line centres were determined by Gaussian fitting. Several lines were
identified on the 2D images and global dispersion coefficients were derived by comparing these lines
with the ThAr line list. Dispersion coefficients for each order were calculated from the global
coefficients and used to make a polynomial approximation to the wavelength scale for each order.

The extracted orders were resampled at constant wavelength intervals and merged into a 1D spectrum
for further analysis.

\begin{figure}
	\centering
		\includegraphics[height=0.47\textwidth,angle=90]{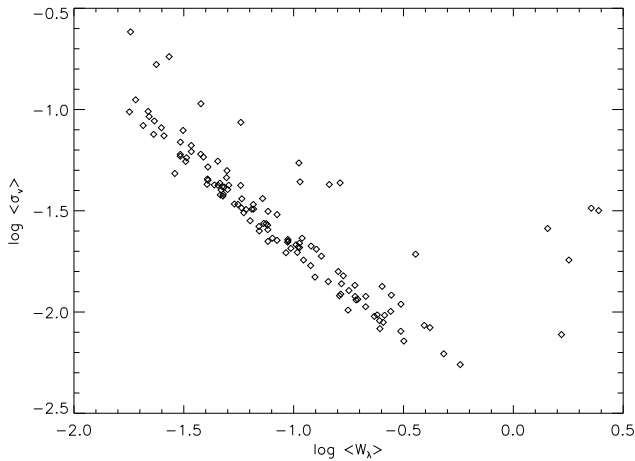}
\caption{The median formal error in the CoG radial velocity errors  $\langle \sigma_v \rangle$ as a function of the median equivalent width $\langle W_{\lambda} \rangle$ for each line.
 }
	\label{f:rverr}
\end{figure}

\section[]{Velocity Measurement}

\subsection{Line measurements}

We first identified a number of lines covering a broad range of atomic species, multiple ions for
the same species and a wide range of excitation potentials and oscillator strengths for given ions.
For each of these we required a radial velocity, equivalent width and associated errors.

For each line, we identified a segment of spectrum (wavelength interval) that covers the single line
in all phases. This segment was extracted from each spectrum and a smoothed version of each segment
was formed using a median filter. The smoothed spectrum was used to estimate the maximum depth in
the line and hence to make a preliminary estimate of the line centre. The centre of gravity (CoG) of
the line profile was obtained from the non-smoothed line segments, using a smaller segment of the
profile with a fixed width centred on the preliminary line centre. The adopted widths are different
for different lines and were estimated iteratively and interactively to obtain a measurement of the
whole profile at all phases. This approach was used to obtain the line wavelengths and also the
equivalent widths. The barycentric radial velocity for each line in each spectrum was obtained by
comparing the CoG wavelength with the laboratory wavelength for that line and correcting for the
Earth's motion.
 
The measurements of weak lines are less reliable, especially when lines are broadened by a large
expansion speed or effective surface gravity around minimum radius, where some lines become quite
asymmetric or simply too weak to measure. Problems may also arise when there are strong nearby lines
or blends and the measurement algorithm selects the wrong component. Such problems can be mitigated
by using a smaller wavelength window, but this has the disadvantage of not including the whole line
profile.

From the original line list covering 13 atomic species and 17 different ions identified in the
spectrum of V652\,Her, we obtained 50\,952 radial velocity measurements covering a total of 132
lines from each of 386 Subaru spectra. Equivalent widths were measured from all 386 Subaru spectra
for 125 lines, the smaller number being due to the dilution of some lines around minimum radius, and
to blending.

The errors in the radial velocity measurements were obtained from the CoG measurement assuming a
mean variance in each datum entering the calculation. For the most part, the mean formal error for
each line was inversely correlated with the strength of the line (Fig.~\ref{f:rverr}):
\[
\log  \langle \sigma_v \rangle \approx - 2 - ( \log \langle W_{\lambda} \rangle + 0.7 ) .
\]
In all cases, the mean errors were $\langle \sigma_v \rangle < 0.2 \kmsec$ for each line and overall $\langle \sigma_v \rangle = 0.06 \kmsec$. 

The radial velocity and equivalent measurements are available in a single table 
as an on-line supplement described in Appendix~\ref{s:app2}.

\subsection{Ephemeris}
\label{s_ephem}

To make further progress with verification and analysis of the line measurements, it was necessary
to convert the times to pulsation phase. The times of mid-exposure were obtained from the image
headers and supplied as geocentric modified Julian date. These were converted to barycentric Julian
dates (BJD) using the FK5 (J2000) coordinates for V652\,Her. We adopted the following ephemeris for
times of maximum based on all times of maxima up to and including 2013 March (Kilkenny 2013,
 private communication):
\[ T_{\rm max} = T_0 + n P_0 + n^2 k_1 + n^3 k_2 + n^4 k_3\]
where $n$ is the cycle number and 
\[\begin{split}
T_0 & = {\rm BJD} 2442216.80481, \\
P_0 & = 0.107992526~{\rm d}, \\
k_1 & = -44.533\times10^{-10}\, {\rm d}, \\
k_2 & = +3.203\times10^{-15}\,{\rm d}, \\
k_3 & = -3.919\times10^{-21}\,{\rm d}.
\end{split}
\]
We note that the period in 2011 had decreased to 0.106994\,d compared with 0.107995\,d originally measured in 1974 by \citet{landolt75}.

\begin{figure}
	\centering
		\includegraphics[height=0.47\textwidth,angle=90]{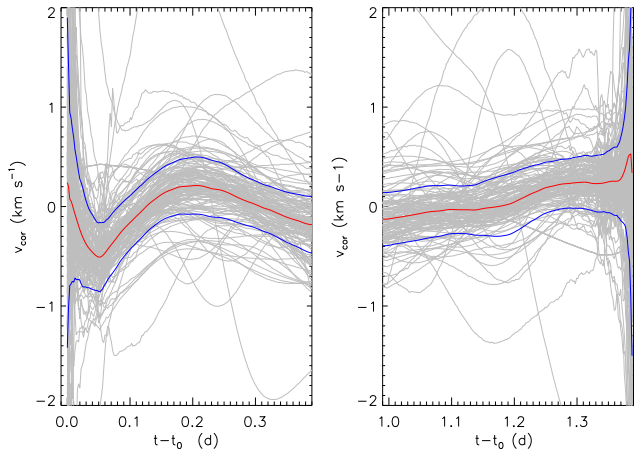}
\caption{The time dependent component of the systematic velocity correction applied to all radial
velocity measurements (solid red line). This correction is evaluated as the smoothed median of the
individual line velocity differences (grey curves) measured relative to a ``representative''
velocity curve. The blue curves show the median curve $\pm$1 standard deviation. The nights of 
June 6 and 7 are plotted separately in the left- and right-hand panels. The most likely causes for these
systematic shifts are temperature changes in the spectrograph and guiding errors on the fibre
bundle.
 }
	\label{f:rvcorrs}
\end{figure}

\begin{figure*}
	\centering
		\includegraphics[height=0.8\textwidth,angle=90]{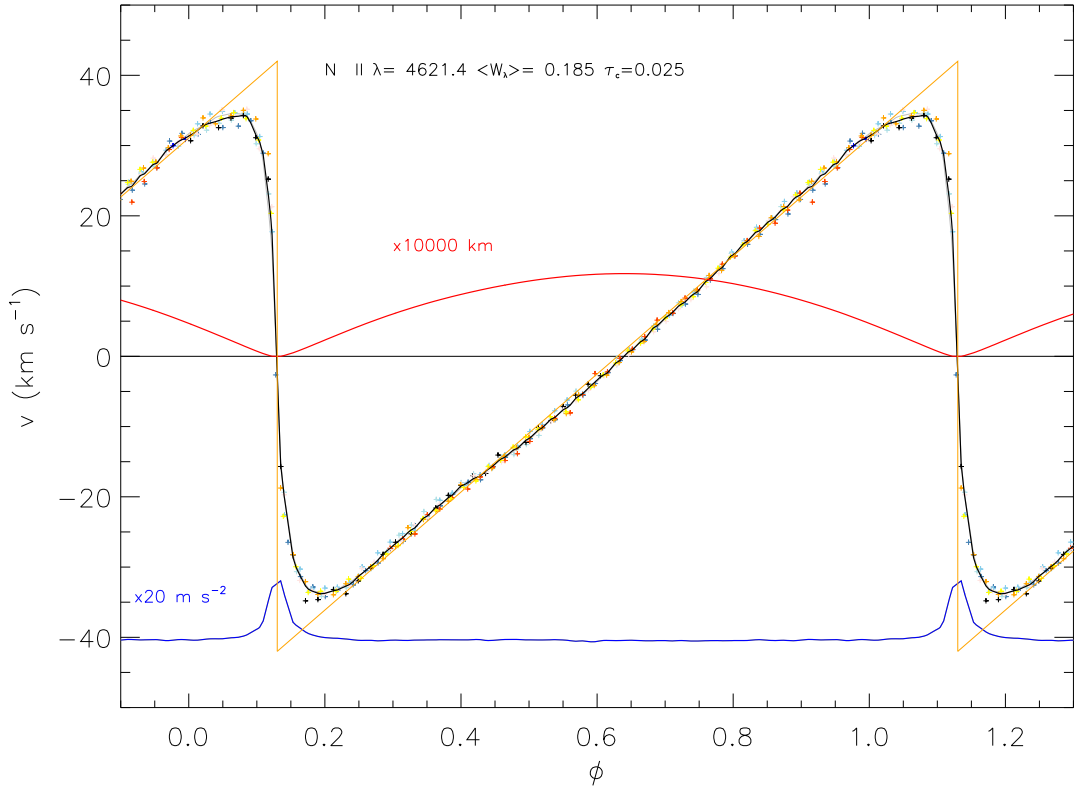}
\caption{
Absorption line velocities and related quantities as a function of pulsation phase for a single
absorption line: in this case N{\sc ii} 4621~\AA. The line is identified by ion and wavelength,
$\lambda$ (in \AA). In addition, the mean equivalent width of the line, $W_{\lambda}$ (in \AA), and
the optical depth of formation of the line core, $\tau_{\rm c}$, are shown. The thin solid {\it
grey} curve represents the reference velocity curve obtained from the median of 29 N{\sc ii} lines.
The thin straight {\it orange} lines are simply guides and represent infinite acceleration followed
by uniform deceleration. {\it Each '+'} represents an individual radial velocity measurement, each
colour refers to a single pulsation cycle, the measurement errors are smaller than the symbols.
Radial velocity: the solid {\it black} curve represents the phase-averaged velocity curve, $v$,
obtained from the line velocities (\kmsec). Not shown: the de-projected surface expansion velocity
$\dot{r}$ is obtained by correcting each line velocity by the appropriate factor (see text).
Acceleration: the solid {\it blue} curve represents the first derivative of the phase averaged
expansion velocity curve, $\ddot{r}$, offset by $-40$ and scaled $\times 20$. Thus, an acceleration
reaching $-30$ on the $y$-axis corresponds to an acceleration of 200 m\,s$^{-2}$. Displacement: the
solid {\it red} line represents the integral of the phase-averaged expansion velocity curve, $\delta
r$, with a constant chosen such that the total displacement over one pulsation cycle is zero and, in
this case, that minimum radius corresponds to a displacement of zero. A displacement of $+10$ on the
$y$-axis corresponds to a radial displacement of 100\,000 km.
 }
	\label{f:line}
\end{figure*}

\begin{figure*}
	\centering
		\includegraphics[height=0.99\textwidth,angle=90]{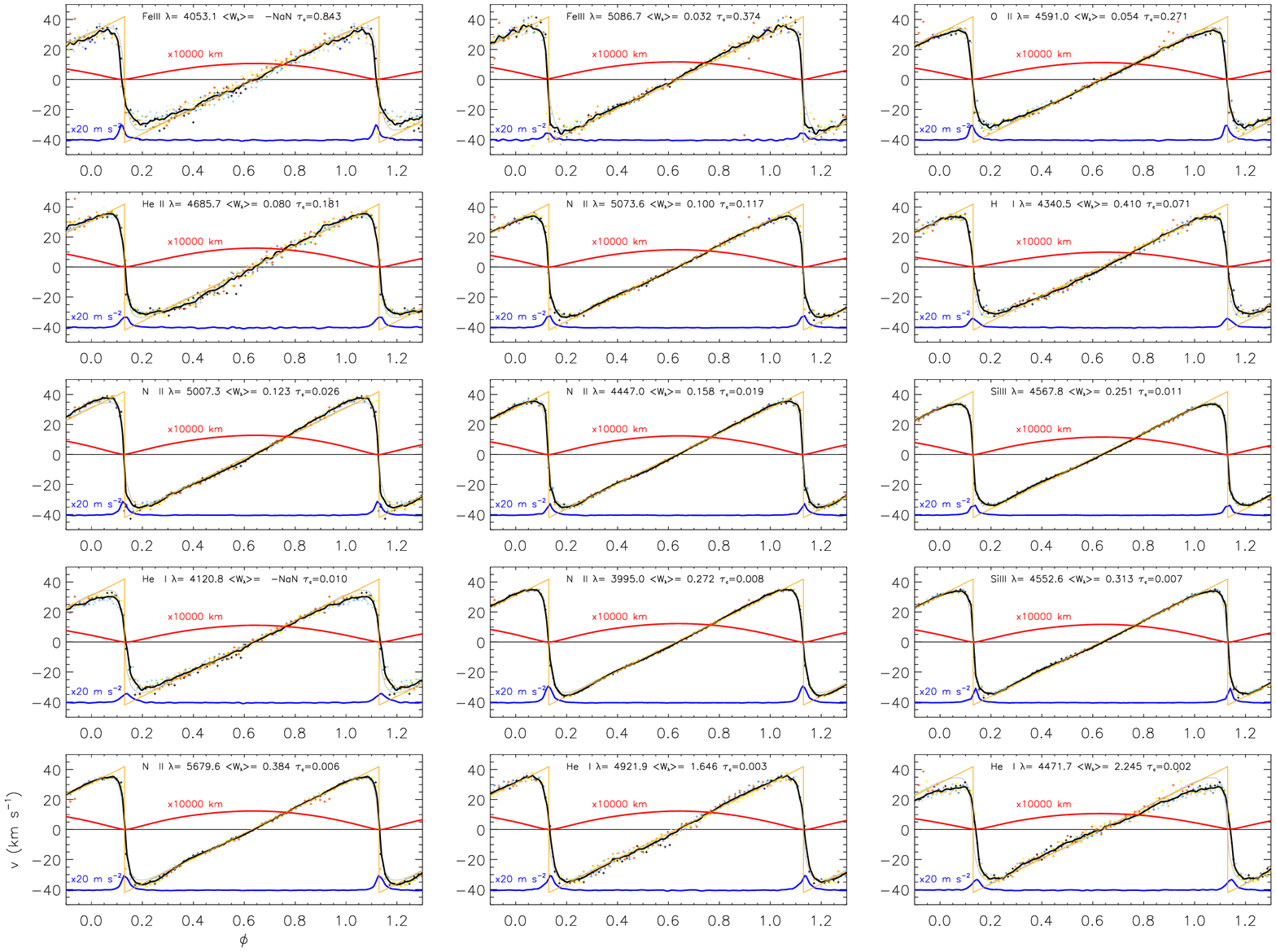}
\caption{As Fig.~\ref{f:line} for an ensemble of absorption lines representing a range of species
and strengths as labelled in each panel. A mean equivalent width $\langle W_{\lambda}\rangle = {\rm
-NaN}$ implies that a valid measurement was not obtained for that line. The phase shift of maximum
acceleration with line depth is best seen by expanding this plot using a machine-readable version of
this paper and a suitable reader.
 }
	\label{f:multi}
\end{figure*}

\subsection{Systematic errors }

Comparing the radial velocity measurements from one line to another, and both radial velocity and
equivalent width measurements from one pulsation cycle to the next, it was clear that two kinds of
systematic error were present: constant and time dependent.

i) Plotting different lines as a function of phase, it was evident that the rest wavelengths used in
each case were not consistent with one another. Since rest wavelengths were not always well
specified, and in many cases lines are blended, these inconsistencies were intelligible and easily
neutralized by the expedient of requiring the mean velocity averaged over all 386 measurements to be
zero.

ii) Plotting velocities and equivalent widths as a function of time showed significant systematic
shifts from one pulsation cycle to another leading to, for example, an apparent overall shift in the
systemic velocity of up to $2\kmsec$ over two nights and a change in the average absorption line
strength by a few per cent. Possible causes for the velocity shifts include motions due to an unseen
companion and/or shifts in the wavelength calibration due to thermal fluctuations in the
spectrograph and/or errors in guiding the star on the slit or fibre bundle. Causes for the
equivalent width shift include inconsistency in continuum placement and/or background (sky,
scattered light) subtraction. We have reduced, but not completely removed these systematic effects
as follows.

To understand the slowly varying velocity shifts we first formed a mean radial velocity curve from a
single strong line (N{\sc ii} 3995~\AA) averaged in phase over all cycles. We subtracted this mean
curve from the observed velocity curve in the time domain, and examined the residual. During the
first part of the first night, the residual velocity dropped evenly by 0.5\kmsec\ over 1 or 2\,h, 
recovered sharply and remained approximately constant for the remainder of the night. On the
second night, the residual increased by about 1.5\kmsec\ during the course of the night. We computed
the same residual for all other lines. Apart from the weakest and noisiest lines, all lines showed
similar behaviour.

Therefore, we established a phase-averaged velocity curve for each absorption line, subtracted this
from the time domain velocity curve for each line, and smoothed the result over 1.2 (FWHM) pulsation
cycles to give a slowly varying velocity correction for each set of radial velocity measurements
(Fig.~\ref{f:rvcorrs}: grey curves). We computed a median and standard deviation from these
corrections, smoothed the result over 0.12 (FWHM) cycles in the time domain and used this correction
for all lines (Fig.~\ref{f:rvcorrs}: red and blue curves). The maximum amplitude of this correction
was $< \pm 0.5 \kmsec$. We investigated the mean residual between these corrected velocities and a
phase-averaged radial velocity curve for each line, and found the mean cycle-to-cycle variation to
lie below $\pm0.4\kmsec$. Increasing the FWHM of the second smoothing function above had a
relatively small effect on the overall results.

Similarly, for the slowly varying equivalent widths we formed a mean equivalent width curve for each
line over a single cycle, and then formed the ratio of this curve with respect to the time varying
curve, smoothed the ratio over 0.5 cycles and applied this slowly varying correction to the original
equivalent width measurements.

\subsection{Radial velocities}

After correction for systematic shifts, all of the radial velocities were plotted as function of
pulsation phase on figures similar to Fig.~\ref{f:line}, a representative plot showing the data for
the relatively strong N{\sc ii} 4621~\AA\ line. The velocities are shown as `+' symbols and colour
coded by pulsation cycle.

The first goal was to determine whether there are line-to-line differences in the radial velocity
curves. In order to assess this, we formed a reference radial velocity curve from the average radial
velocities of 29 strong N{\sc ii} lines for which the lines were strong and present in all
observations. These were ordered by phase and smoothed using a Gaussian filter with FWHM of 0.016
cycles. The radial velocities for each line measured were also ordered by phase, and smoothed using
the same filter function. The reason for smoothing the data was to obtain as optimal a phase
resolution as possible, using observations from successive pulsation cycles obtained at different
phases to fill in the phase space. Limitations are provided by incomplete removal of the systematic
cycle-to-cycle zero-point shifts, and a relatively non-uniform phase distribution of the data,
despite observing nearly six distinct cycles without a formal constraint on the exposure start times.
An experiment with a smaller Gaussian (0.012 cycles) resulted in increased errors resulting from the
non-uniform phase distribution

We examined the difference in the sense of observed-minus-reference velocity for each line. In the
case that an absorption line enters the rapid acceleration phase earlier than the reference line,
the residual will show a trough centred at or before the point of maximum acceleration. If the
absorption line accelerates later than the reference line, the residual will show a peak after
maximum acceleration. Examples of both types of behaviour were readily identified.

The hypothesis is that deeper layers of the photosphere commence acceleration before higher layers
as the “shock” propagates upwards through the photosphere. The assumption was that higher ionization
species (e.g. Fe{\sc iii}, Si{\sc iv}) would be formed deeper than lower ionization species (e.g.
Si{\sc ii}, N{\sc ii}). In fact, there were no systematic differences among Si{\sc ii}, {\sc iii}
and {\sc iv}. What became clear was a strong systematic correlation between line {\it strength}
(expressed as equivalent width) and acceleration phase.

This can be understood as follows. The central depth of an absorption line is a direct measurement
of the position in the photosphere at which the line optical depth $\tau_{\lambda}$ is unity. In
continuum regions, photons arrive from regions where the continuum optical depth $\tau_{\rm c}<1$.
In line centres, photons arrive from layers where $\tau_{\lambda}<1$. If the atmosphere structure
and composition are known, the relation between geometric depth and line optical depth can be
computed. Alternatively, one can write the position at which $\tau_{\lambda}=1$ in terms of some
continuum optical depth $\tau_{\rm c}$. For now, one only needs to recall that the deeper an
absorption line, the higher in the photosphere that the core is formed, the limit being the layer
above which scattering dominates over absorption.

This result is partially illustrated in Fig.~\ref{f:multi}, which presents the same information as
Fig.~\ref{f:line} for 15 representative lines from the very weak Fe{\sc iii} 4053~\AA\ to the very
strong He{\sc i} 4471~\AA.

\begin{figure}
	\centering
		\includegraphics[height=0.47\textwidth,angle=90]{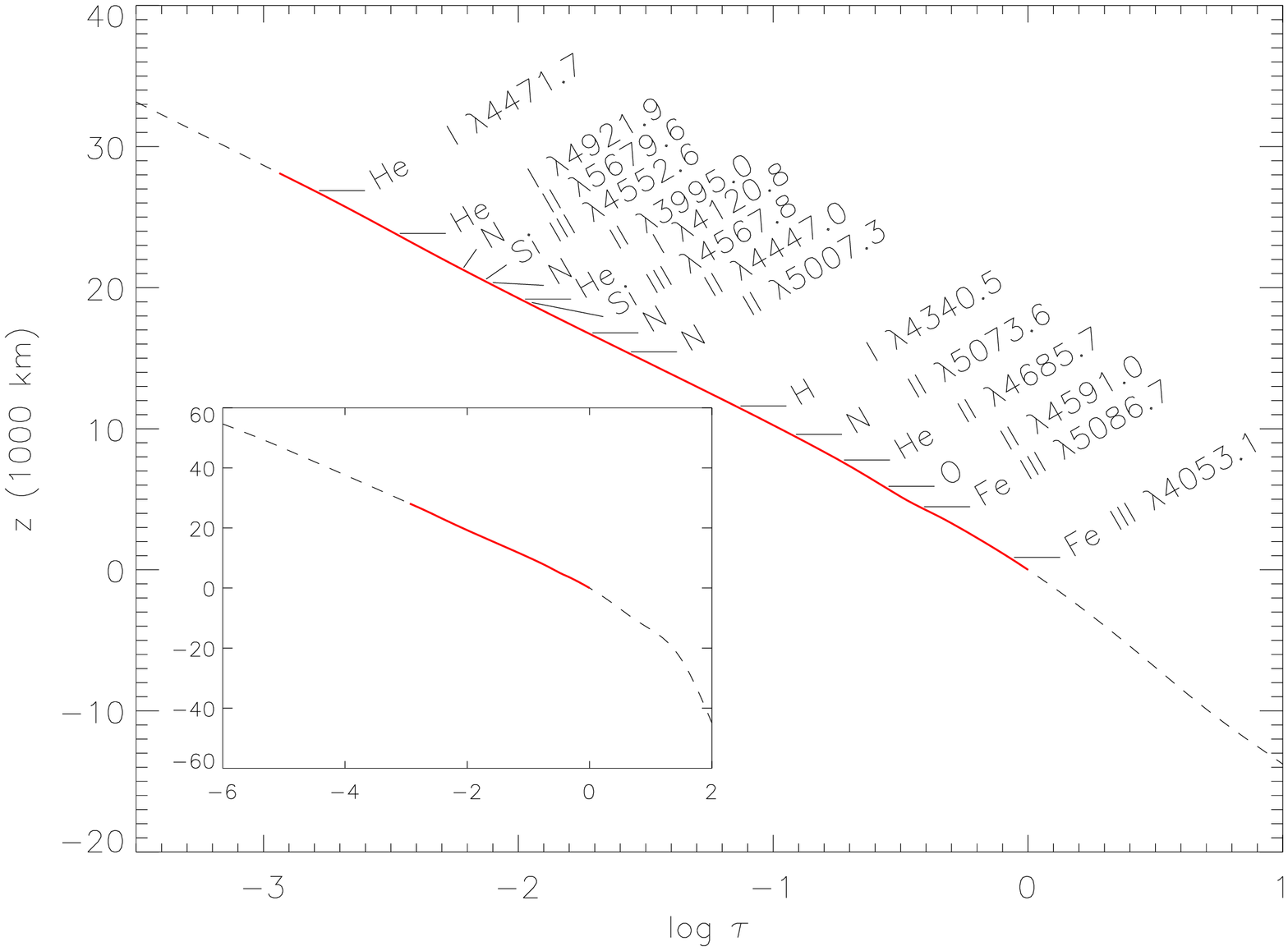}
\caption{Detail from the run of geometric height above the continuum forming layer $z(\tau)-z(1)$ as
a function of optical depth $\tau_{4000}$ in a model for the atmosphere of V652\,Her (broken line).
The line-forming region is marked as a solid red line. The locations of formation of the cores of
the representative lines in Fig.~\ref{f:multi} are indicated. The same function for the entire model
atmosphere is inset. }
	\label{f:lf}
\end{figure}

\section{Pulsation}

To study the behaviour of each line individually and the photosphere represented by the entire ensemble, we have analysed the data as follows.

\subsection{Model atmosphere}
\label{s:atmos}

To establish where in the photosphere the core of each line is formed, we computed a representative
model for the atmosphere of V652\,Her \citep{behara06} and its emergent spectrum \citep{jeffery01b}.
This model was selected using a median Subaru spectrum obtained between phases 0.60 and 0.70, and
without any velocity corrections. This spectrum, and the adopted model are shown in
Appendix~\ref{s:app1}.

\subsection{Photospheric structure}

The model spectrum includes, for every line, the value of $\tau_{4000}$, the optical depth in the
continuum at which the line optical depth is unity. The model also permits us to compute a
relationship between $\tau_{4000}$ and the geometric depth relative to some reference point. In this
case we adopt the layer where the continuum optical depth at 4000~\AA\ is unity (Fig.~\ref{f:lf}).
Thus the geometric position of the line core is established, assuming an equilibrium structure. For
the present, we posit that, to first order, the structure of the photosphere at maximum radius
matches this structure satisfactorily. For reference, Fig.~\ref{f:lf} identifies the positions at
which the cores of the representative lines shown in Fig.~\ref{f:multi} are formed, showing that
they cover nearly three decades in optical depth, and some 25\,000\,km in geometric depth (at
maximum radius). To place some of these numbers in context, we note that the mean radius of
$2.31\pm0.02\Rsolar$ \citep{jeffery01b} corresponds to $\approx1\,600\,000$ km. Consequently, the
region of the star sampled by the absorption line cores represents $\approx 2\%$ of the radius.

\subsection{Projection effect}

The radial velocity $v$ measured for a spherically expanding star is diluted by projection effects
(sometimes referred to as limb darkening), and so must be increased by a suitably chosen projection
factor $p$ in order to represent the {\it true} expansion velocity of the star $\dot{r}$
\citep{parsons72}. \citet{montanes01} studied this question for the general case of early type
stars, for V652\,Her in particular, and as a function of line strength. Following Parsons, they
adopted a relation for the projection factor of the form
\[ \dot{r} = - p v, \] 
\[  p = a_0 + a_1 \gamma \]
where
\[ \gamma = - \dot{r}/w_{1/2} \]
and 
\[w_{1/2} = \frac{\Delta  \lambda_{1/2}c}{\lambda}\]
is the half width (in velocity units) at half depth of the line at wavelength $\lambda$. For the case of V652\,Her, we adopt  values from \citet{montanes01} (Table 4, Case (b)):  
\[ a_0=1.402, a_1 = -0.028. \]
Values for $\Delta  \lambda_{1/2}$ were obtained from the same model spectrum used to establish the geometric  structure of the photosphere. For $\gamma$, we adopt the first order approximation $\dot{r} = -1.38 v$.

\subsection{Displacement and acceleration}

Having established $\dot{r}$ for each absorption line, it is straightforward to evaluate the local
acceleration $\ddot{r}$ and displacement $\delta r$ by differentiating and integrating,
respectively. The latter requires a boundary condition; we here require the displacement to be zero
at maximum radius. These quantities are illustrated for selected lines in Figs.~\ref{f:line} and
\ref{f:multi}.

The choice of boundary condition allows us to add the geometric depth at which the line forms at
maximum radius, as deduced above, and hence obtain, as a function of phase, the radial displacement
from every line measured relative to a defined position, being where $\tau_{4000} =1$ at maximum
radius. We also adjusted the zero-point of the velocity curve (a second time) to ensure that the
total displacement over the pulsation cycle was zero.

\begin{figure}
	\centering
		\includegraphics[height=0.47\textwidth,angle=90]{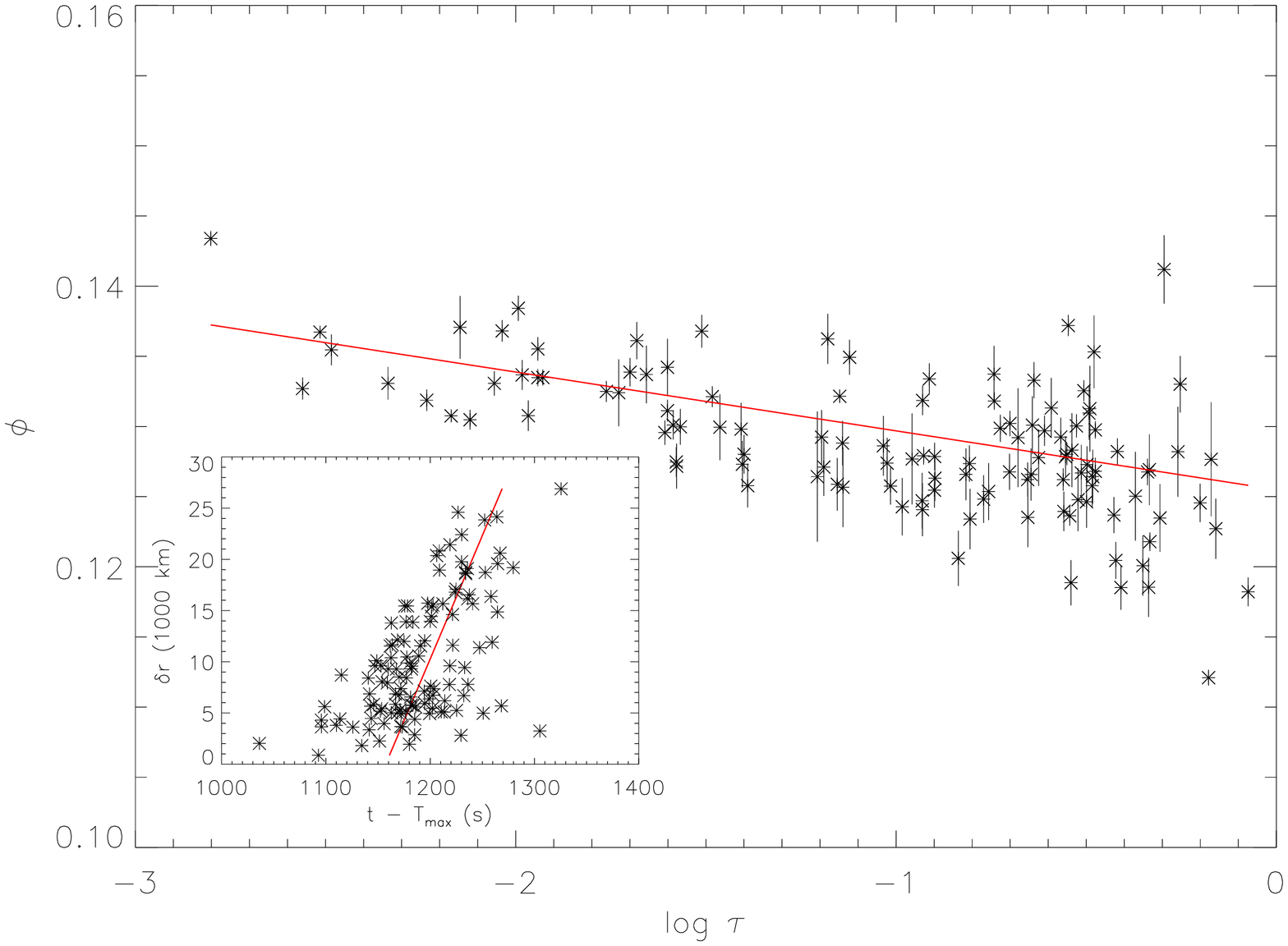}
\caption{The phases of maximum acceleration obtained from 137 absorption lines as a function of optical depth of  formation of the line cores. The linear fit to these data { (red line) and $1\sigma$ errors  are } also shown. The inset panel shows the same information plotted as a function of geometric depth against time since $T_{\rm max}$.    }
	\label{f:accel}
\end{figure}

\begin{figure}
	\centering
		\includegraphics[height=0.47\textwidth,angle=90]{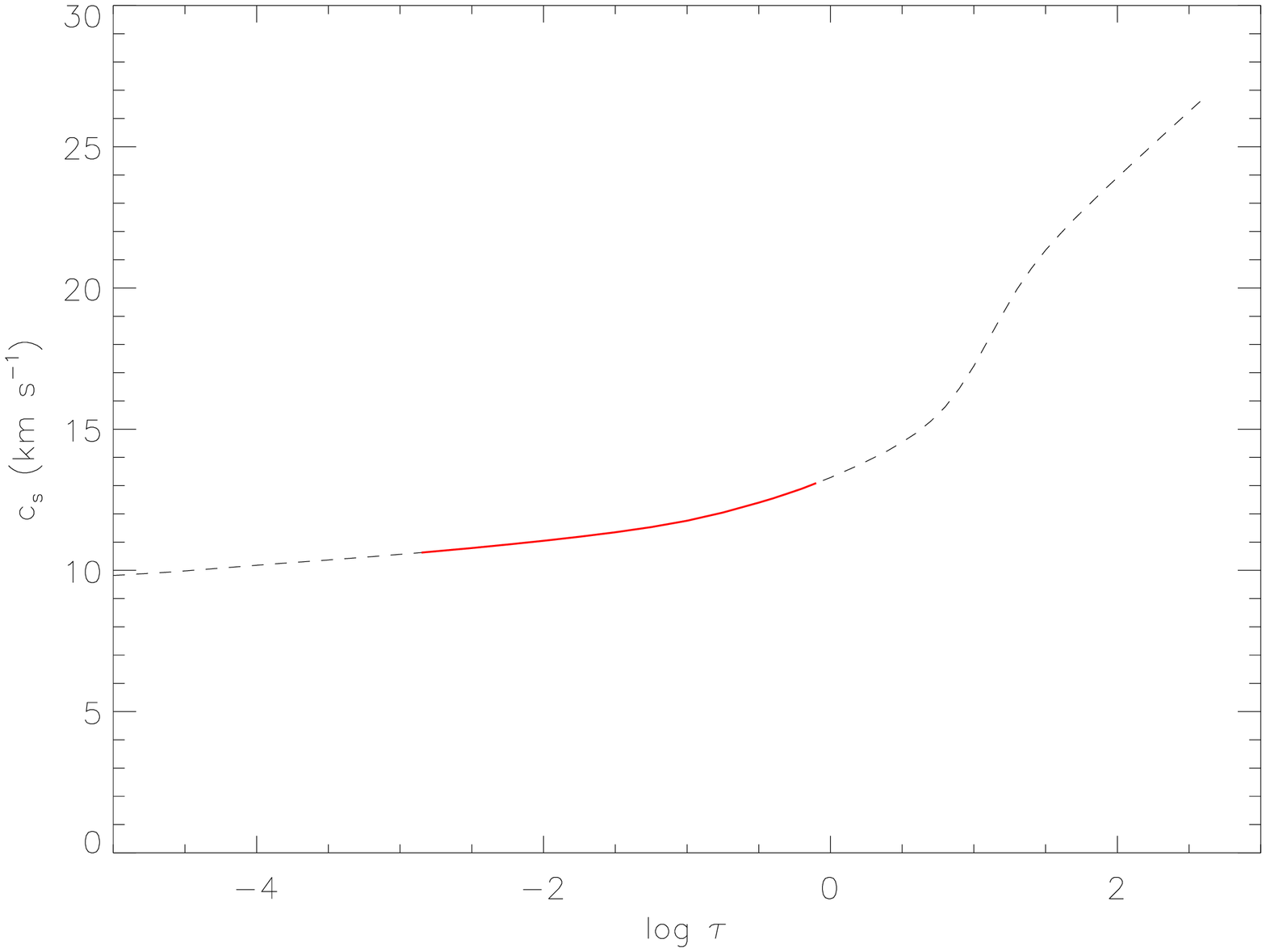}
\caption{The local sound speed $c_{\rm s}$ in a hydrostatic LTE model atmosphere for V652\,Her at maximum radius 
as a function of continuum optical depth (broken line). The line-forming region (Fig.\,\ref{f:lf}) is shown as a solid red line. 
}
	\label{f:ss}
\end{figure}

\subsection{Phase of minimum radius}

The phase of minimum radius for each line was measured from the  phase of maximum acceleration
$\phi_{\ddot{r}_{\rm max}}$. We located the maximum by fitting a parabola to the acceleration curve,
including up to 20 points within $\pm0.05$ cycles of the largest value of $\ddot{r}$. The results
are shown in Fig.~\ref{f:accel} as a function of the line core optical depths. Although there is
substantial scatter at large optical depths, where the absorption lines are weakest and hardest to
measure, a trend is apparent. A linear fit giving the correlation, \[ \phi_{\ddot{r}_{\rm max}} =
0.1254(2) - 0.0042(1) \log \tau \] is also shown in Fig.~\ref{f:accel} (errors on final digits in
parenthesis). This demonstrates that maximum acceleration, and hence minimum radius, occurs earlier
in deeper layers, and supports the hypothesis stated in \S4.4. The outward moving pressure wave
emanating from the stellar interior that characterises the pulsation subsequently propagates upwards
through the photosphere. Since we have a good estimate of the geometric depth in terms of optical
depth (Fig.~\ref{f:lf}), the above correlation converts to a mean pulse speed through the
photosphere $v_{\rm pulse} = 239\pm6 \kmsec$ (Fig.\,\ref{f:accel} inset). For reference, the local
sound speed in a stationary atmosphere for the model adopted to represent the star at maximum radius
is shown in Fig.\,\ref{f:ss}.
 
\begin{figure}
	\centering
		\includegraphics[height=0.46\textwidth,angle=90]{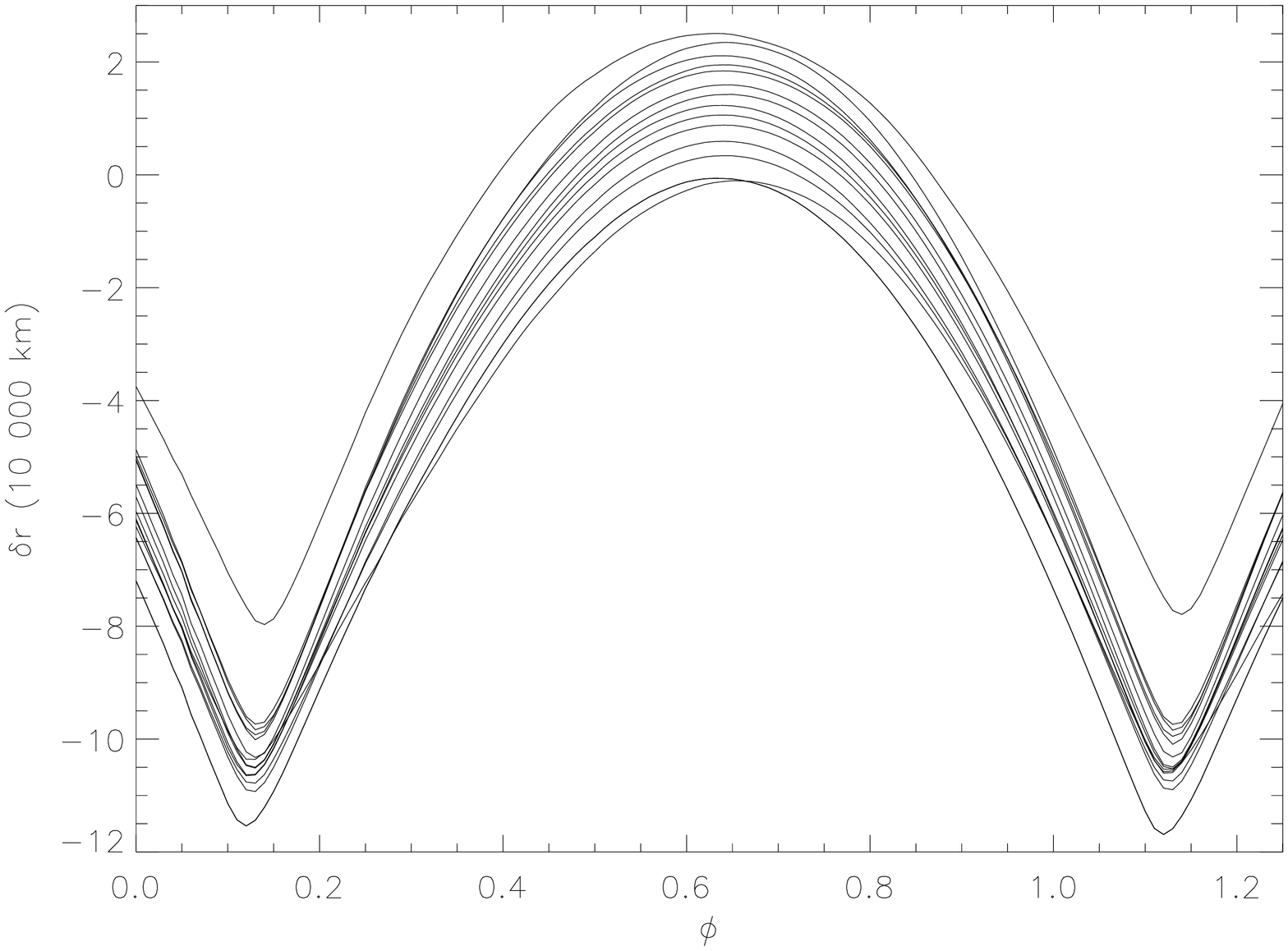}
		\includegraphics[height=0.47\textwidth,angle=90]{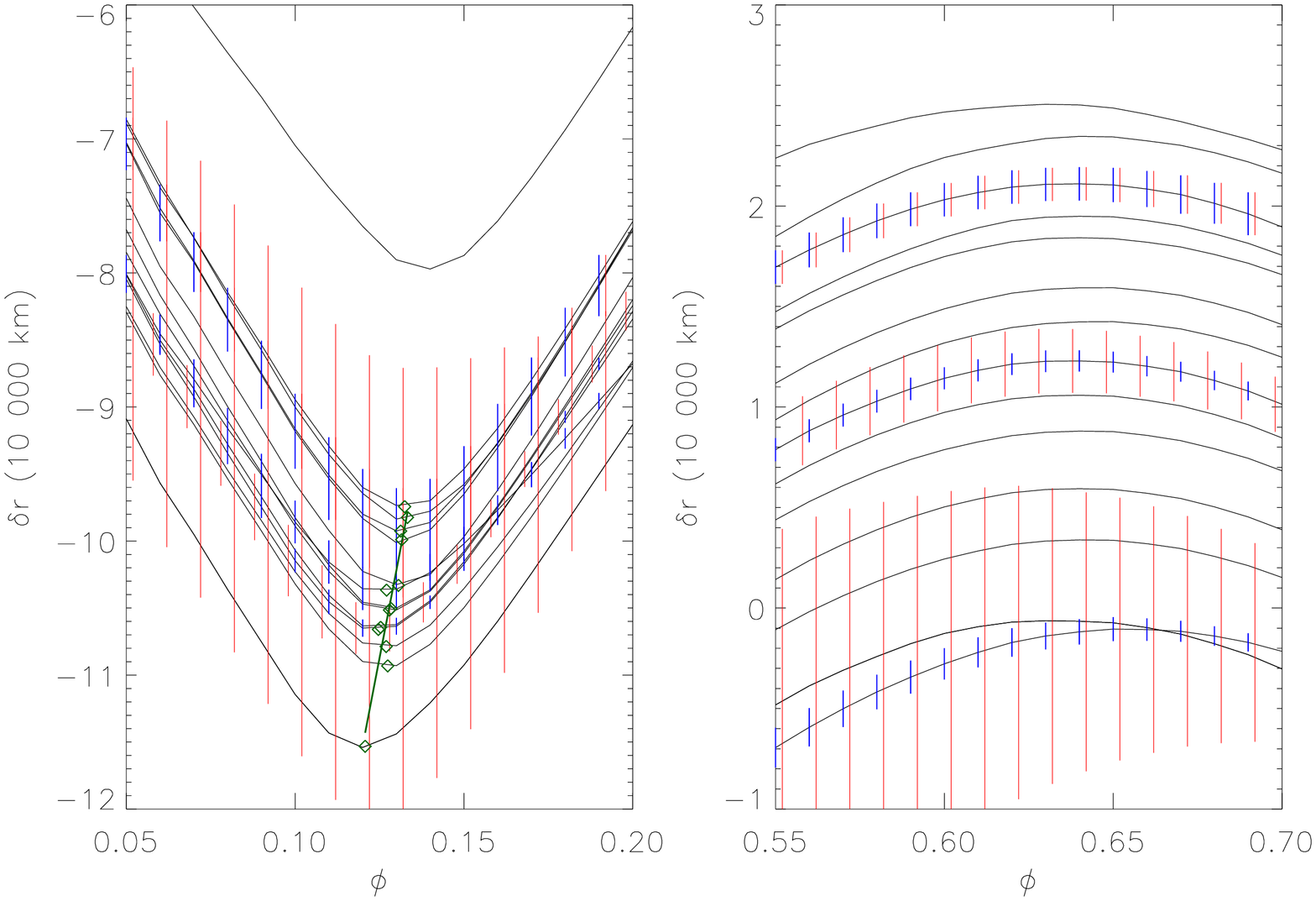}
\caption{Top: the total radial displacement of different layers of the photosphere of V652\,Her as
represented by different groups of absorption lines. The lines are grouped by depth of formation at
maximum radius into bins with $\delta \log \tau_{\rm c}=0.2$ centred at $\log \tau_{\rm c} = -0.1
(0.02) -2.9$, and normalized using the $\tau-z$ relation shown in Fig.~\ref{f:lf} such that a line
formed at $\tau_c=0$ would have zero displacement at maximum radius. Consequently the weakest lines
correspond to the lowest layers of the photosphere and vice versa. Bottom: Expansions of the upper
panel around minimum and maximum radius. Since each curve represents the average of several lines
within a bin, standard deviations (red) and standard errors in the mean (blue) have been added for
three representative bins. The green diamonds represent radius minima for each line group obtained
by fitting a parabola through the five points around minimum radius; the solid green line represents
a regression on these minima (excluding He{\sc i} 4471\AA) with a slope ${\rm d} \delta r /{\rm
d}\phi \times {\rm d}\phi/{\rm d}t=141\pm17\kmsec$.}
	\label{f:displace}
\end{figure}

\subsection{Resolving vertical motion}

Another way of representing the pulsation of the photosphere is by considering the radial
displacement. The individual line displacements obtained by integrating $\dot{r}$ and adding the
geometric depth of formation can be plotted as a function of phase. Owing to noise in the weaker
lines, a plot showing all individual lines is not informative. By grouping lines in terms of
$\tau_c$ and plotting the median displacement for each group, the time varying structure of the
photosphere becomes apparent. We used uniform bin sizes $\delta \log \tau_c=0.2$ centred at $\log
\tau_c=-0.1, -0.3, \ldots, -2.9$. Fig.~\ref{f:displace} demonstrates how the photosphere is
compressed by almost a factor of two as the star approaches minimum radius, and then recovers as the
star expands. Again placing numbers in context, the mean pulsation amplitude of $\approx120\,000$ km
represents $\approx8\%$ of the mean stellar radius. The top layer is represented by a single very
strong line, He{\sc i}~4471~\AA, for which radial velocities are difficult to measure. The bottom
layers are represented by extremely weak lines which virtually disappear at minimum radius. However,
the overall result holds well at all depths, and also reflects the phase dependence of minimum
radius obtained using $\ddot{r}$ above. By fitting parabolae to each line group, an alternative
estimate of the phase of radius minimum is obtained (Fig.~\ref{f:displace}: lower left). A
regression on these locations gives the vertical velocity of minimum radius as $141\pm17\kmsec$.
Whilst this number is nearly a factor of 2 smaller than that obtained from the phases of maximum
acceleration, the manner in which the displacements were obtained is significantly different.

It should be emphasized that the motion depicted in Fig.~\ref{f:displace} is {\it not} Lagrangian.
The displacements traced here refer to the locations of formation of the cores of specific groups of
absorption lines. These locations move both {\it with} the bulk motion of the photosphere {\it and
within} the photosphere as it is compressed and heated. In order to understand Fig.~\ref{f:displace}
in terms of the motion of mass, a theoretical model of the spectrum derived from an appropriate
hydrodynamical model of the pulsating star will be necessary.

Whether the phase of maximum acceleration or of minimum radius is used, one thing is clear: the
passage of either quantity through the photosphere is of the order of 10 times the value of the
local sound speed.

\subsection{He {\sc i} 4471~\AA}

He {\sc i} 4471~\AA\ is difficult to measure for two major reasons. First, it is very strong, so the
CoG method for velocity measurement requires a very broad window if it is to be consistent with
other lines. However, this means that other lines become included in the blend. Second, the triplet
and forbidden components, which form at different depths, are strongly blended and hence the line is
asymmetric; moreover, due to the quadratic Stark effect and electron impacts, both the widths and
rest wavelengths of the lines shift with density \citep{barnard69}. Consequently, in the context of
pulsation, the line does not have a fixed rest wavelength, as was assumed for the velocity
measurement. To some extent, other diffuse lines including He {\sc i} 4921~\AA\ and 4026~\AA\ suffer
from the same problem.
 
\subsection{Surface trajectory}

In \S\,\ref{s:intro}, V652\,Her was introduced as `the rocket star' because of its characteristic
pulse-coast-pulse shaped radial velocity curve. Since the `coasting' phase appears almost linear,
one might ask how closely the surface motion follows a ballistic trajectory. Applying the projection
factor described above, the mean acceleration $\langle \ddot{r} \rangle = -12.1\pm0.4 ~{\rm
m\,s^{-2}}$ between $0.3 < \phi < 1.0$ over all \ion{N}{ii} lines. It is not exactly linear
(Figs.\,\ref{f:line}, \ref{f:multi}), but increases in amplitude from $-10.9$ to $-12.5 {\rm
m\,s^{-2}}$ between the first and the second halves of this interval.

For comparison, the gravitational acceleration of a star with $R=2.3\Rsolar$, and $M=0.59\Msolar$ is
twice this value with $g = -25 ~{\rm m\,s^{-2}}$ (corresponding to $\log g=3.4$ in cgs units). Since
the radial displacement (120\,000 km) is not small compared with the radius, the true surface
gravity varies between $-26 ~{\rm m\,s^{-2}}$ at minimum radius, and $-23 ~{\rm m\,s^{-2}}$ at
maximum. A ballistic trajectory would be symmetric around maximum displacement.

Thus the observed surface trajectory is not truly ballistic, and the increasing deceleration is
presumably due to a more gradual decay in internal pressure following the major pulse at minimum
radius. If the surface trajectory were to be closer to ballistic, the effective gravity acting on
the photosphere would be much reduced, with significant consequences for the spectrum. \section{Line
Profile Variations}

The principal advantages of using one radial velocity measurement per absorption line per spectrum
are the limitation in the number of data and the relative simplicity of interpretation. A
disadvantage of the CoG approach is that additional information provided by the whole line profile
is lost. At large radii, the near ballistic atmosphere should show narrow absorption lines
characteristic of low surface gravity whilst, around minimum radius, the accelerating atmosphere
should show broader lines characteristic of high surface gravity. In between times, the lines are
intrinsically asymmetric due to projection effects from the spherical surface. In addition, around
minimum radius and where conditions permit, there is the possibility of observing both
inward-falling and outward-rising material in the same absorption line. This would be indicative of
highly compressed and possibly shocked material. This section seeks to interpret the line profile
variations over the pulsation cycle of V652\,Her.

\begin{figure*}
	\centering
		\includegraphics[width=0.90\textwidth,angle=0]{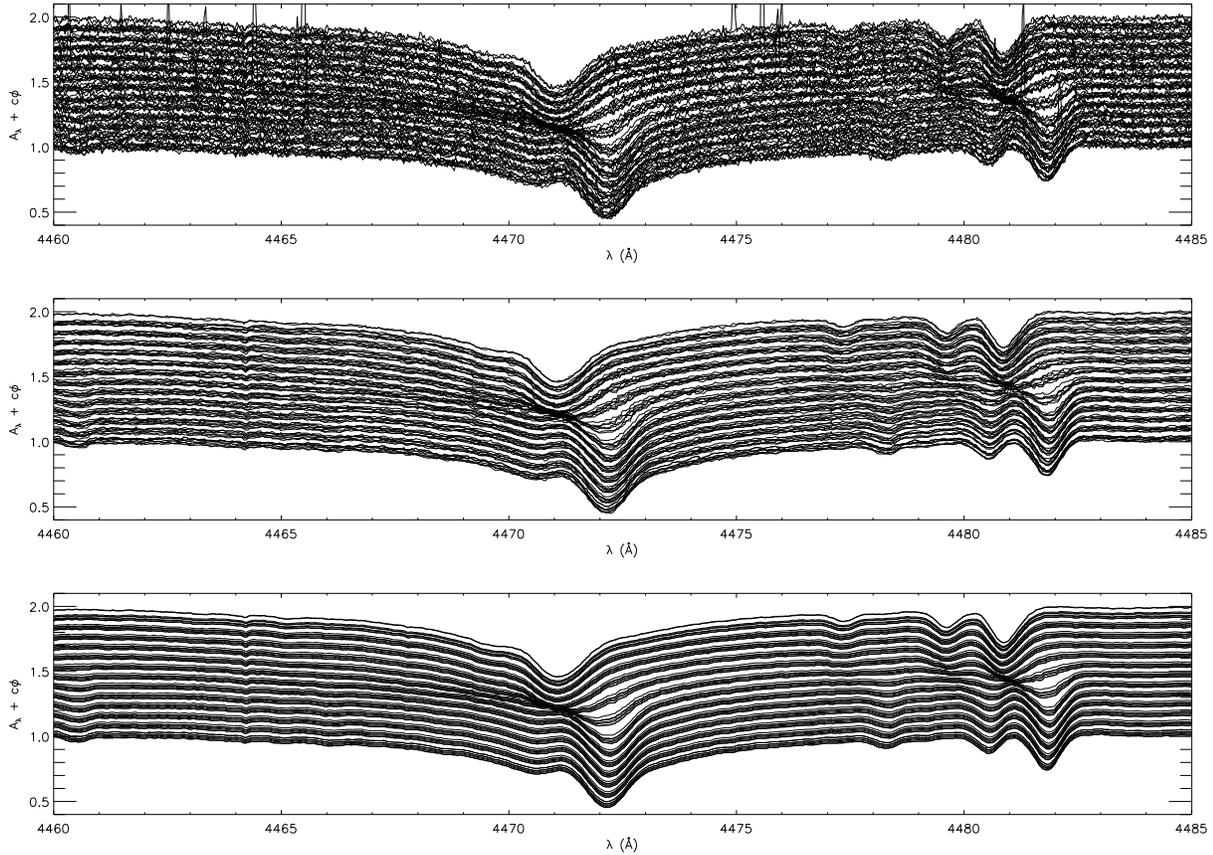}
\caption{The phase folded spectrum of V652\,Her around the  He{\sc i} 4471~\AA\ and Mg{\sc ii} 4482~\AA\ lines and
around minimum radius. The top panel shows the individual normalized spectra ($A_{\lambda}$), offset vertically upwards by an amount proportional to phase $\phi$. The middle panel shows the same data after application of a 4-pixel median filter in the dispersion direction. The bottom panel shows the data after application of a Gaussian filter  in the phase direction with FWHM = 0.024 cycles.  }
	\label{f:4471_rmin}
\end{figure*}

\subsection{Optimizing signal-to-noise ratio}

Although the S/N ratio of most individual spectra is high, the number of spectra obtained close to
minimum radius is relatively small, the exposure time relative to the acceleration phase is
significant, and the effects of gravity broadening and acceleration smearing dilute information at
these phases. However, there are two major ways in which the S/N ratio can be further increased.
First, the resolution of the Subaru data (90\,000) is sufficiently high that the intrinsic profiles
of even the weakest lines are oversampled. Secondly, by effectively observing eight passages through
minimum radius over the two nights we can increase the S/N by folding the data in phase. One
approach would be to bin the data in both wavelength and phase; we have rejected this to avoid
losing information unwittingly. Instead, we have applied a 4-pixel median filter in the dispersion
direction and then applied a Gaussian filter with an FWHM of 0.012 cycles in the phase direction. The
result is demonstrated for the He{\sc i} 4471~\AA\ and Mg{\sc ii} 4482~\AA\ lines around minimum
radius in Fig.~\ref{f:4471_rmin}. Since the Gaussian filter has an FWHM comparable with the duration
of a single exposure (0.013 cycles), little information is lost\footnote{The Gaussian filter is one
half that used for phase averaging the radial velocities. Tests showed that noise reduction with the
smaller filter was nearly as good as with the larger, and better preserves the phase resolution.}.

\begin{figure}
	\centering
		\includegraphics[width=0.47\textwidth,angle=0]{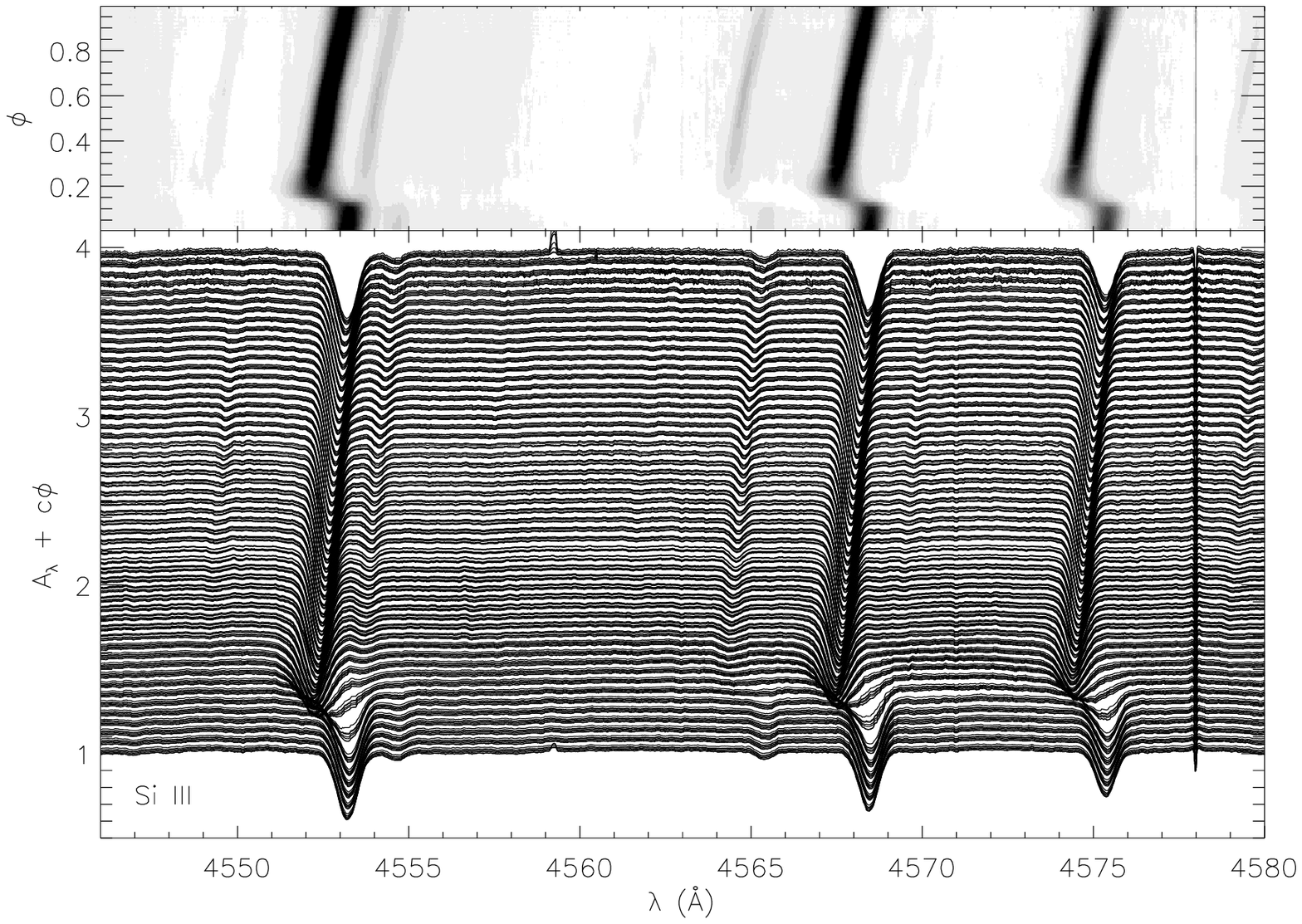}
                  \includegraphics[width=0.47\textwidth,angle=0]{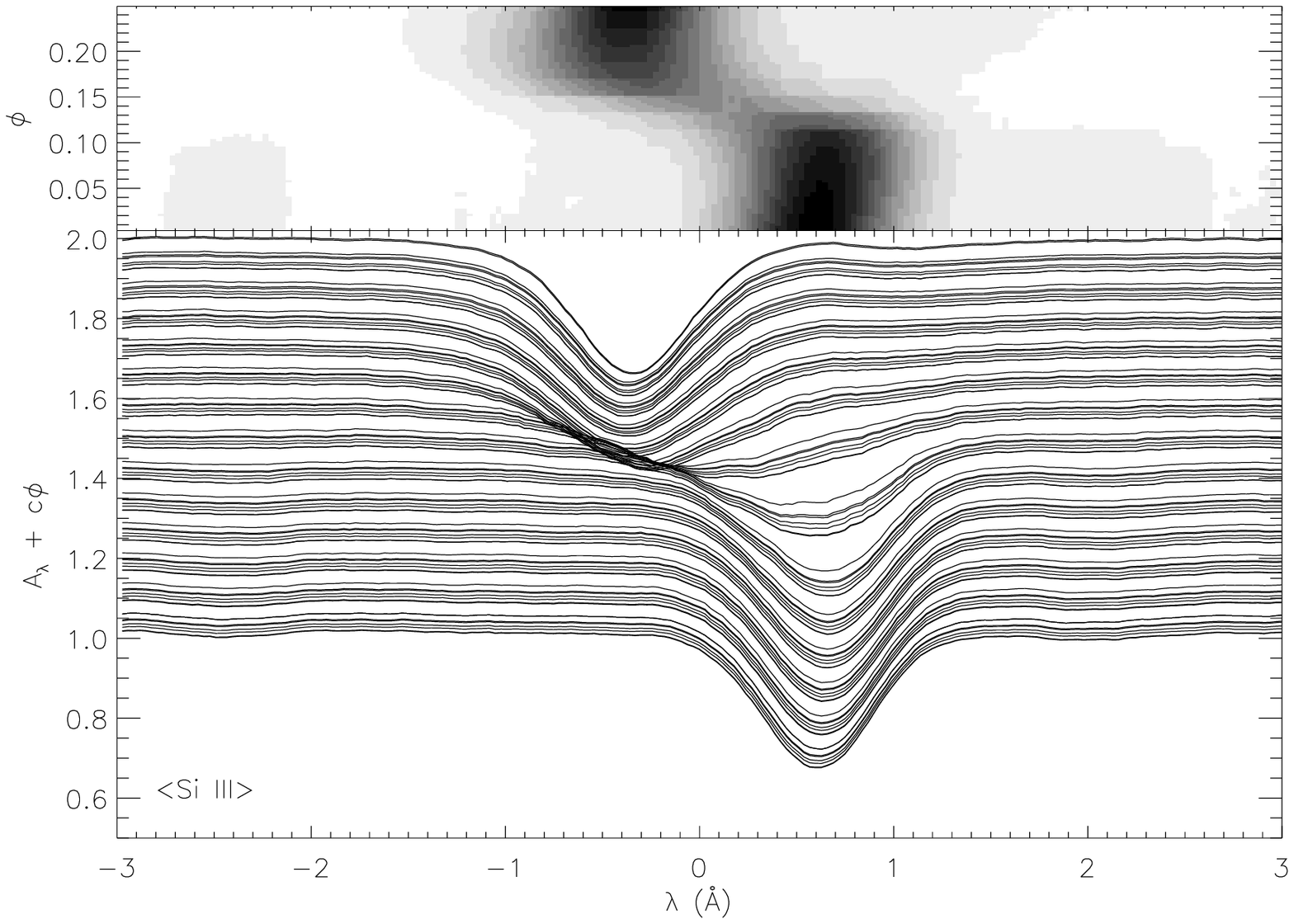}
\caption{The behaviour of the strong Si{\sc iii} triplet over a complete pulsation cycle (top) and around minimum radius (bottom). In each pane, the lower panel represents the continuum normalized flux as a  function of wavelength, with a constant added which is proportional to the phase. The upper panel represents a grey-scale plot containing the same information. Features which do not change in wavelength as a function of phase are instrumental artefacts. The bottom panel represents the mean of all three Si{\sc iii} lines. The wavelength scale is relative to the rest wavelength for each line. The data have been phase smoothed (see Fig.~\ref{f:4471_rmin}).  }
	\label{f:si3}
\end{figure}

\begin{figure}
	\centering
		\includegraphics[width=0.47\textwidth,angle=0]{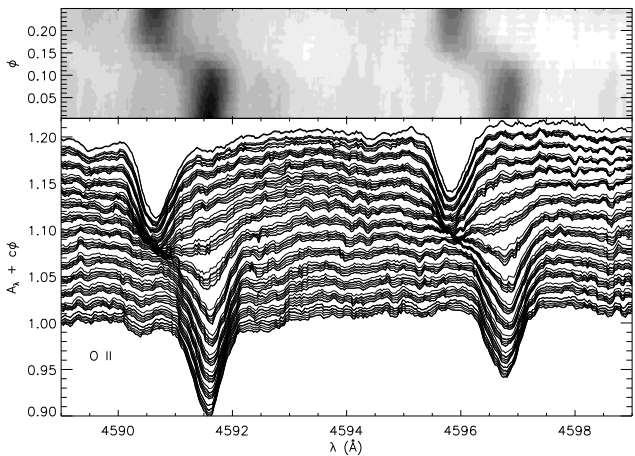}
\caption{The behaviour of  two intermediate strength O{\sc ii} lines around minimum radius. The data have been phase smoothed  (see Fig.~\ref{f:4471_rmin}). }
	\label{f:o2}
\end{figure}

\begin{figure}
	\centering
		\includegraphics[width=0.47\textwidth,angle=0]{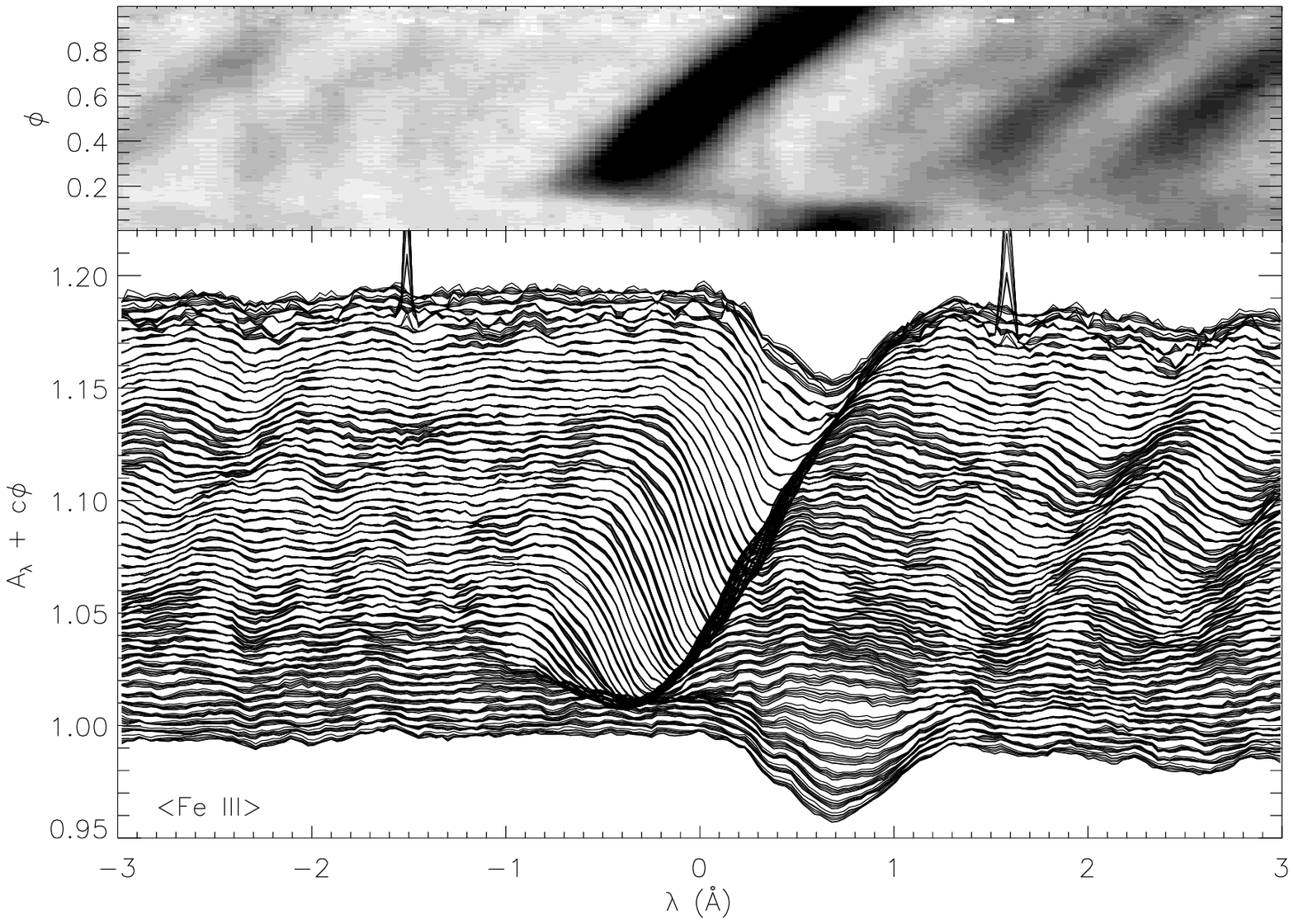}
		\includegraphics[width=0.47\textwidth,angle=0]{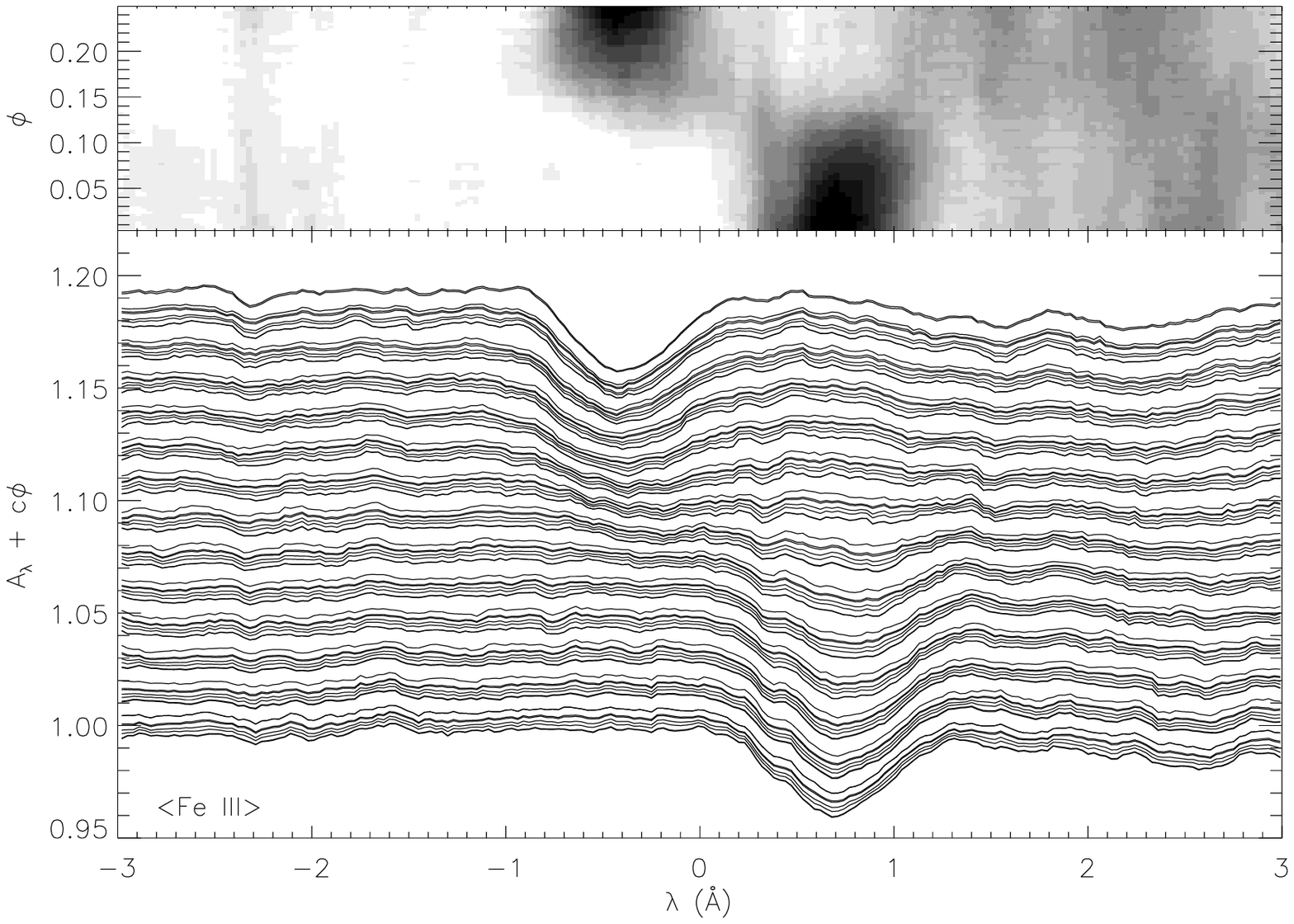}
\caption{The behaviour of six weak  Fe{\sc iii} lines  represented as a composite line profile relative to the line rest wavelength. The wavelength scale is relative to the rest wavelength for each line. The data have been phase smoothed (see Fig.~\ref{f:4471_rmin}). }
	\label{f:fe3}
\end{figure}

\subsection{Line behaviour}

Figs~\ref{f:si3} - \ref{f:fe3} illustrate the behaviour of a number of individual absorption
lines over the whole pulsation cycle and expanded over the phase around minimum radius. In the cases
of strong lines, this behaviour is relatively easy to identify. The two obvious features are (a)
narrowing of the lines from phases 0.3 to 0.65, followed by broadening, and (b) the sharpness of the
transition from contraction to expansion. Qualitatively speaking, the transition appears to be
almost continuous in the case of the Si{\sc iii} lines (Fig.\,\ref{f:si3}). In the case of the
weaker O{\sc ii} lines (Fig.\,\ref{f:o2}), there appears to be an almost complete break. Although
the CoG of the line appears to shift smoothly (see Fig.\,\ref{f:multi}), its depth
becomes much shallower.

Three questions are important. (1) Does the absorption shift smoothly in velocity, or does the red
shifted component simply disappear to be replaced by a new blue shifted component? (2) If the latter
is true, what is the physics behind it? (3) Are both components ever visible as independent
absorption lines in the same spectrum?

In order to investigate lines deeper in the photosphere, the weakest lines must be studied, but even
with the quality of data available, noise is still significant. In addition to wavelength and phase
smoothing, we have combined data from several similar absorption lines into a single composite
profile by coadding in wavelength space at the rest wavelength of each line (Fig.\,\ref{f:fe3}).

Fig.~\ref{f:fe3} combines data for six Fe{\sc iii} lines (5193.9, 5235.7, 5282.3, 5299.9, 5302.6,
and 5460.8~\AA). In this representation, the red shifted and blue shifted components become almost
completely detached at minimum radius.

On the basis of these data, there does appear to be a jump discontinuity in the position of the weak
absorption lines. The red-shifted component formed in downward moving material is effectively
replaced by a blue shifted component formed in upward moving material. The transition takes place
within 0.01\% of a pulsation cycle (about 150\,s, or roughly the time resolution of the
observations).

The physics is harder to explain; as material moves downwards it compresses and heats; it also meets
upward moving material which is denser and hotter, but which is expanding and cooling. For the line
{\it not} to move smoothly from red to blue is counter-intuitive, unless it is associated with the
passage of some kind of shock or discontinuity through the line-forming medium. 

\section{Conclusions}

We have presented {\it Swift} ultraviolet photometry and Subaru HDS spectroscopy of the pulsating
helium star V652\,Her. The {\it Swift} photometry provides the best definition of the ultraviolet
light maximum so far, and the best relative photometry. The data also emphasize a need for higher
quality UV observations of this star. No evidence was found for a shock at ultraviolet or X-ray
wavelengths.

The Subaru HDS spectroscopy offers high spectral and high temporal resolution data over six pulsation
cycles (and eight radius minima). The temporal resolution (176\,s) is sufficient to sample the rapid
acceleration phase well, although some velocity smearing is evident. Observations obtained through
several radius minima allow some improvement to this resolution. The data have enabled a
line-by-line analysis of the entire pulsation cycle and provided a description of the pulsating
photosphere as a function of optical depth, demonstrating how it is compressed by a factor of at
least 2 at minimum radius. This analysis also demonstrates that the phase of radius minimum is a
function of optical depth and allows the speed of the pulse running upwards through the photosphere
to be measured at between 141 and 239 \kmsec, depending how it is measured. This speed is at least
10 times the local sound speed, implying that it must generate a shock wave as it passes. The
strong acceleration at minimum radius is demonstrated in individual line profiles; those formed
deepest in the photosphere show a jump discontinuity of over 70\kmsec\ on a times-cale of 150\,s,
providing further evidence of a shock.

The next step in the analysis of these data will be to generate a good hydrodynamic model of the
pulsations. Such a model will be strongly constrained by the pulsation period, by the observed motion
(amplitude and acceleration) of the surface, and by the ultraviolet light amplitude presented in
this paper. One objective of these calculations must be to explain the origin of the phase lag of
+0.15 cycles between maximum light and minimum radius in V652\,Her. Phase lags exist in Cepheids but
in the opposite sense (--0.25 cycles) and are understood in terms of linear (non-adiabatic) effects
associated with the thin hydrogen ionization zone moving through mass layers almost as a
discontinuity \citep{castor68,szabo07}. Since the hydrogen and first helium ionization zones play no
role in V652\,Her (too little hydrogen and too hot), it is suggested that { an interaction
between the second helium ionization zone and the nickel/iron opacity bump at $2\times10^5$\,K will
be important.}

The final step will be to couple the pulsation model to the observations presented here to determine
more precisely the overall properties of V652\,Her and, in particular, its mass. \citet{jeffery01b}
assumed a quasi-static approximation to measure effective temperature and effective surface gravity
throughout the pulsation cycle using models in hydrostatic equilibrium. This approximation can be
removed by coupling the hydrodynamical pulsation model to a radiative transfer code, with which we
will be able to simulate the Subaru data realistically.

\section*{Acknowledgements}

The authors are grateful to  Matt Burleigh for procuring the NGTS photometry, and
to Wayne Landsman, Mat Page, Sam Oates and Paul Kuin for helpful discussions regarding the UVOT data.
They are also grateful to Hideyuki Saio for continuing encouragement,  discussion and helpful 
observations on the manuscript, and to the referee Giuseppe Bono for his useful remarks. 

The Armagh Observatory is funded by direct grant from the Northern Ireland Department of Culture, Arts and Leisure.
RLCS is supported by a Royal Society fellowship. 

This work was partially carried out with support from a Royal Society UK-Japan International Joint Program grant to DK and HS.

\bibliographystyle{mn2e}
\bibliography{ehe}

\appendix
\renewcommand\thefigure{A.\arabic{figure}} 
\renewcommand\thetable{A.\arabic{table}} 

\section[]{Line radial velocity and equivalent width measurements}
\label{s:app2}

\begin{table*}
\caption{Times of observation, line identifications, radial velocities and equivalent widths for
the  observations described in \S4.1. Times and velocities are 
barycentric. The format is decribed within the data file at the start of each section: 
  '//' represents a linefeed separating scalars and vectors.  Lines commencing  '\# ' are descriptive comments. 
 In this extract, the first few lines of each 
section are reproduced; ellipses represent omitted lines. }
\label{t:lines}
{\sffamily \flushleft 
\# V652 Her - Subaru 2011 - Radial velocity and equivalent width measurements\\
\# Raw measurements (no systematic corrections) \\
\# --------------------------------------------------------------------------\\
\# barycentric julian dates of midexposure - 2450000 \\
\# nt // times(0:nt-1) \\
         386 \\
  5719.23633  5719.23828  5719.24023  5719.24219  5719.24414  5719.24658 \\
... \\
\# -------------------------------------------------------------------------\\
\# absorption lines identified by central wavelength, atomic number and ion\\
\# nl // lambda(0:nl-1) //  atom(0:nl-1) // ion(0:nl-1) :  \\
\# atom = 0 : line measured but not retained \\
\# ion = 0 : neutral, 1 : singly-ionized, ... \\
     139 \\
 3994.997 4009.258 4026.191 4035.081 4039.160 4041.310 4043.532 4053.112 4056.907
...\\
  7  2  2  7 26  7  7 26  7
...\\
  1  0  0  1  2  1  1  2  1
...\\
\# --------------------------------------------------------------------------\\
\# barycentric radial velocities and formal errors, identified by wavelength\\
\# n, lambda(n) // v(0:nt-1,n) // sigma\_v(0:nt-1,n) : [ km/s ]\\
\# n $<$ 0 : block terminator \\
    0  3994.997 \\
  34.770  35.609  36.487  38.518  37.178  37.526  38.090  31.245   7.133
...\\
   0.024   0.016   0.017   0.014   0.012   0.010   0.011   0.005   0.012
...\\
    1  4009.258 \\
...\\
 -1 0  \\
\# --------------------------------------------------------------------------\\
\# equivalent widths and formal errors, identified by wavelength\\
\# n, lambda(n) // W\_lambda(0:nt-1,n) : [ AA ]\\
\# n $<$ 0 : block terminator \\
    0  3994.997 \\
 0.240 0.243 0.245 0.238 0.247 0.248 0.249 0.247 0.262 0.252 0.255 0.257
...\\
    1  4009.258 \\
 1.441 1.439 1.408 1.428 1.368 1.391 1.382 1.370 1.321 1.359 1.323 1.331
...\\
 -1 0  \\
\# --------------------------------------------------------------------------\\
}
\end{table*}

Times of observation, line identifications, radial velocity and equivalent width measurements 
obtained from  the  observations described in \S\S\,3 and 4 are
given in a single ascii file as an on-line supplement. 
 Table\,\ref{t:lines} illustrates the structure of the file, which is divided into four blocks, each one 
representing the four quantities just described. Although 139 lines are formally listed in the
second section and were indeed measured, there are a few instances where the results were
unsatisfactory; these data are not reported. They are indicated by setting the atomic number 
for those lines to zero. They are retained as place holders to maintain the integrity of the
line identification system, given by $n=0,nl-1$ in the subsequent sections. 
Zero-based numbering is  used for computational convenience.

\renewcommand\thefigure{B.\arabic{figure}} 
\renewcommand\thetable{B.\arabic{table}} 

\section[]{The optical spectrum of V652\,Her  at maximum radius}
\label{s:app1}

Figure~\ref{f:atlas_a} (Parts (b)--(h) online only) shows the median spectrum of V652\,Her around maximum radius and between
pulsation phases 0.60 and 0.70, together with a model and identifications by ion for all absorption
lines with theoretical equivalent widths $W_{\lambda}>10$~m\AA. A few lines are missing from the
model. A few lines are stronger in the model than in the observed spectrum. This reflects
incompleteness in the atomic data. The optimized spectrum is obtained from a fully line blanketed
model atmosphere in hydrostatic and local thermodynamic equilibrium. The adopted model has
atmospheric parameters $T_{\rm eff}=22\,500$\,K, $\log g=3.3$ (\cmss), $n_{\rm H}=0.005$, $n_{\rm
He}=0.98$ (number fractions), and $v_{\rm turb}=9 \kmsec$. Other abundances in the model are
$n_{\rm C}=0.00005$,   
$n_{\rm N}=0.0014$,  
$n_{\rm O} =0.0001$,
$n_{\rm Ne}=0.00026$,
$n_{\rm Si}=0.000076$,  and   
$n_{\rm Fe}=0.00003$.   
Given that the atmosphere is certainly not in hydrostatic equilibrium and some ions are
probably not in LTE \citep{przybilla05}, the model is used primarily for line identification 
and to approximate the atmospheric properties. Greater consistency in the wings
of Stark broadened helium lines,  for example, would be desirable.


\begin{figure*}
\centering \includegraphics[width=0.90\textwidth,angle=0]{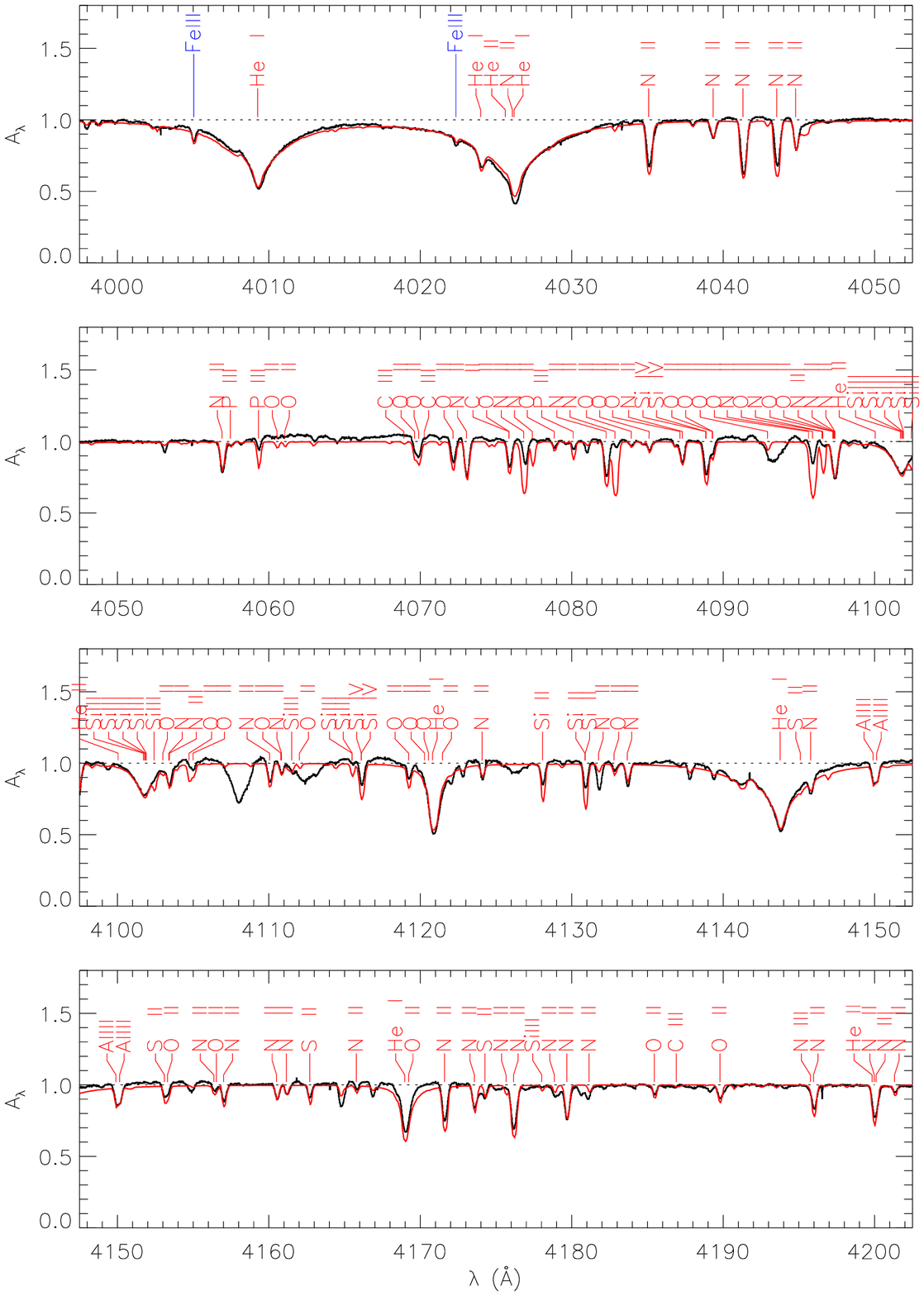}
\caption{(a) The median Subaru spectrum of V652\,Her between phases 0.60 and 0.70 (black), together with an approximate LTE synthetic spectrum computed using the parameters given in \S~\ref{s:atmos} (red). Lines in the model with $W_{\lambda}>10$~m\AA\ are identified by ion. Ions with $Z>20$ are shown in blue. The hydrogen Balmer lines at 4101~\AA\ and 4340~\AA\ are not labelled but are included in the model.  The model continuum is shown as a dotted line. Strong broad features around 4093, 4108, 4112 and 4126~\AA\  are instrumental artefacts.  } \label{f:atlas_a}
\end{figure*} 
\begin{figure*} \addtocounter{figure}{-1}
	\centering
		\includegraphics[width=0.90\textwidth,angle=0]{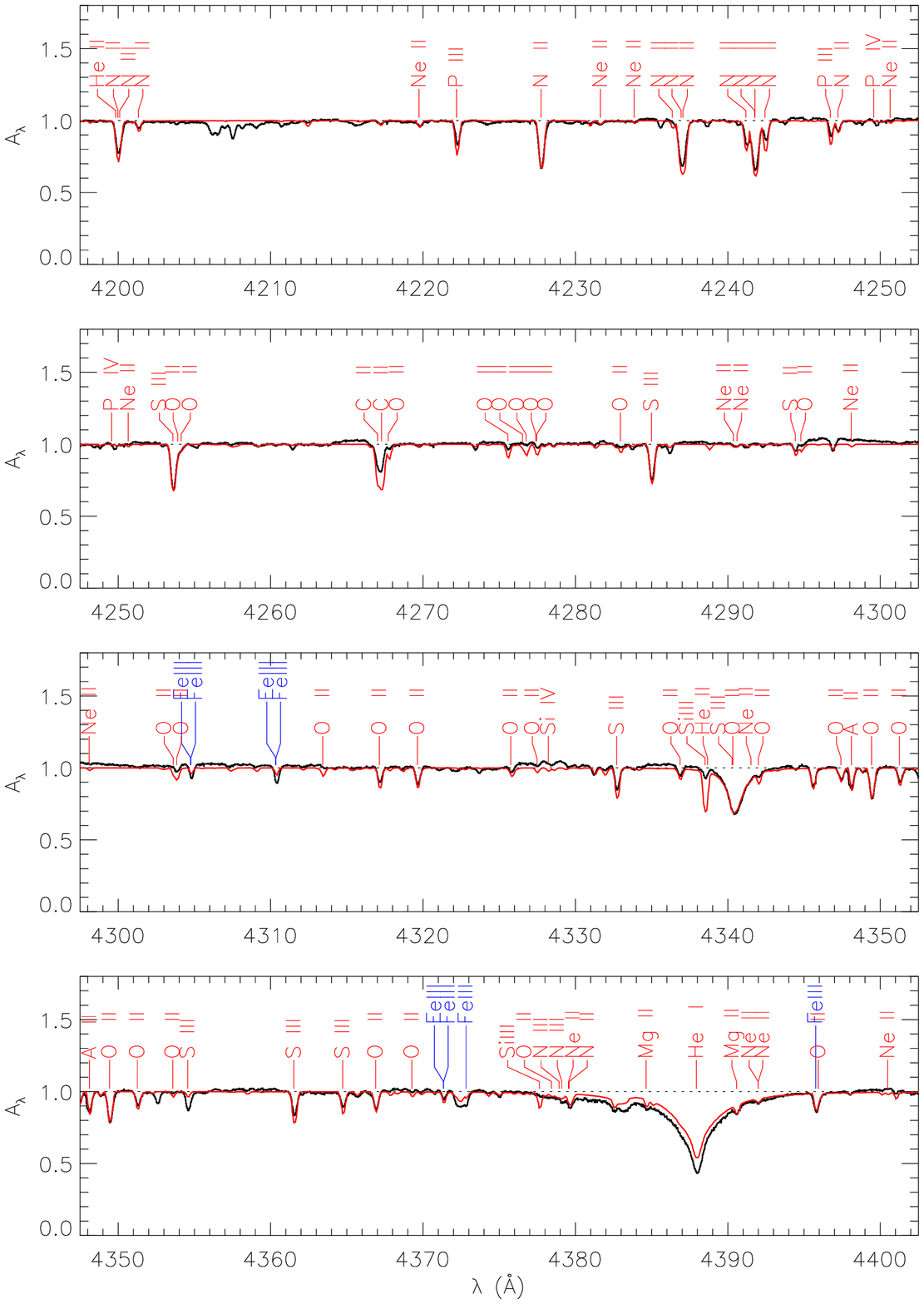}
\caption{(b) contd.  } \label{f:atlas_b}
\end{figure*}   
\begin{figure*}\addtocounter{figure}{-1}
	\centering
		\includegraphics[width=0.90\textwidth,angle=0]{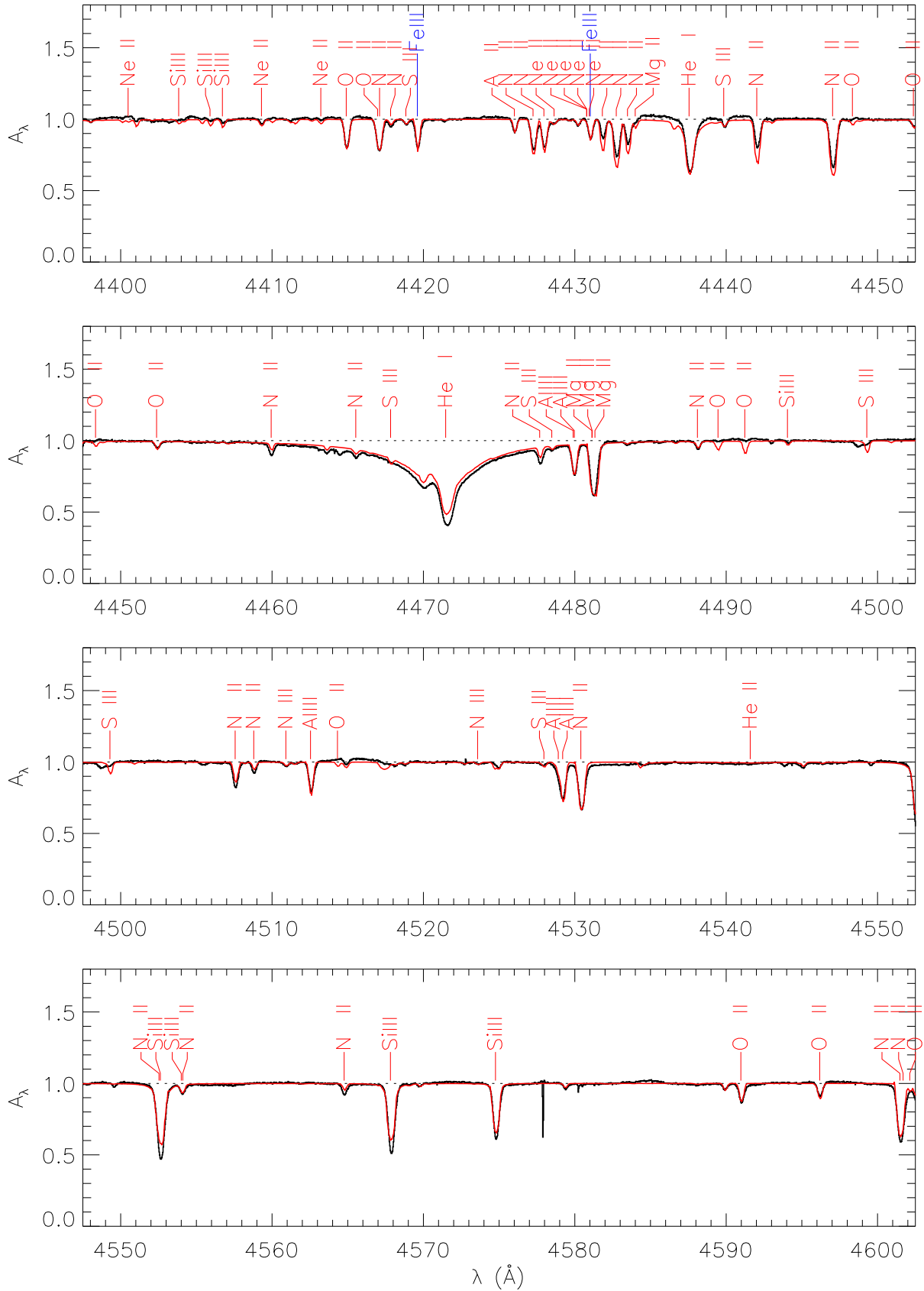}
\caption{(c) contd.  Sharp features around  4578 and 4580~\AA\  are instrumental artefacts } \label{f:atlas_c}
\end{figure*}   
\begin{figure*}\addtocounter{figure}{-1}
	\centering
		\includegraphics[width=0.90\textwidth,angle=0]{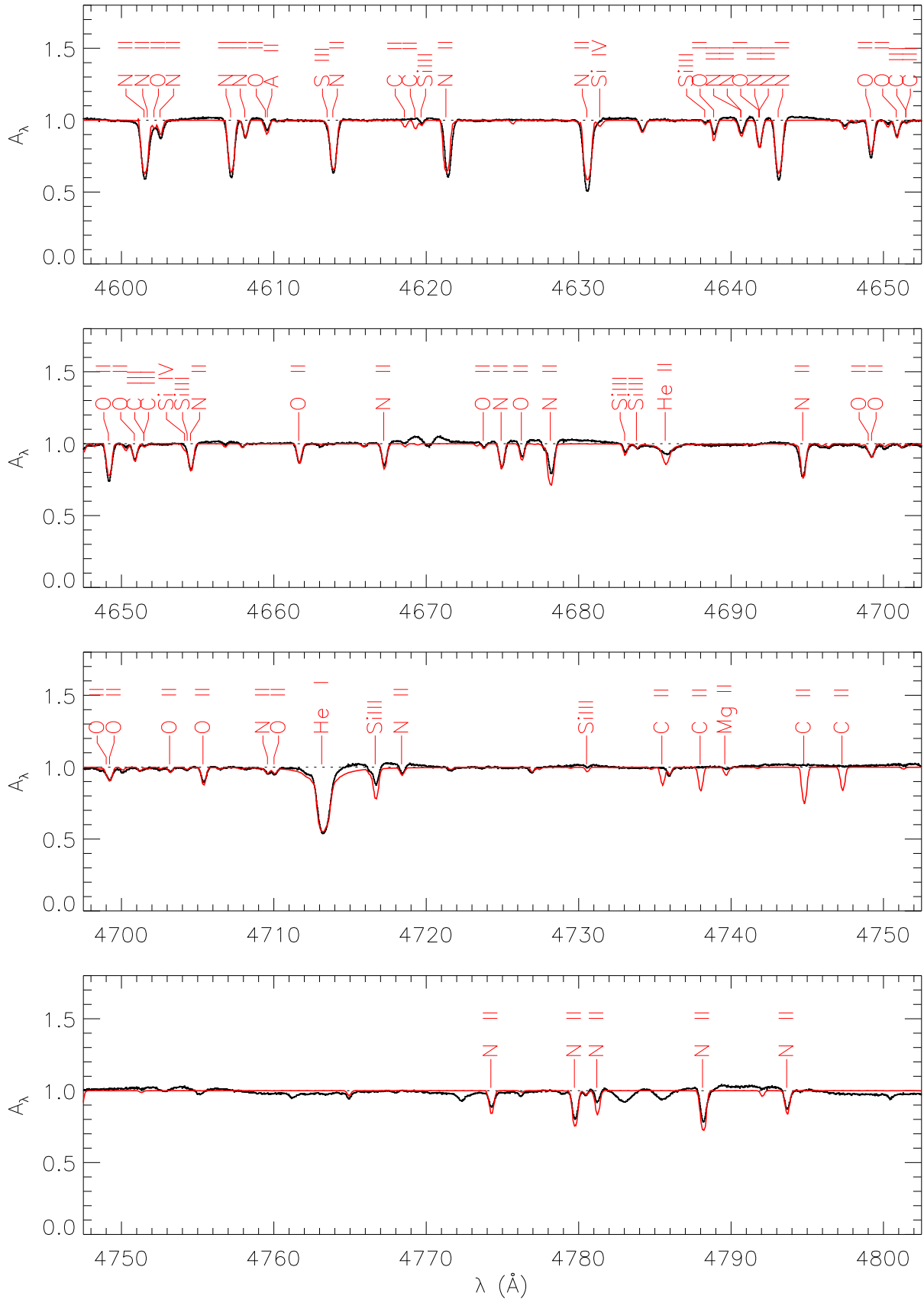}
\caption{(d) contd. Strong broad features around  4783 and 4786~\AA\  are instrumental artefacts.} \label{f:atlas_d}
\end{figure*}   
\begin{figure*}\addtocounter{figure}{-1}
	\centering
		\includegraphics[width=0.90\textwidth,angle=0]{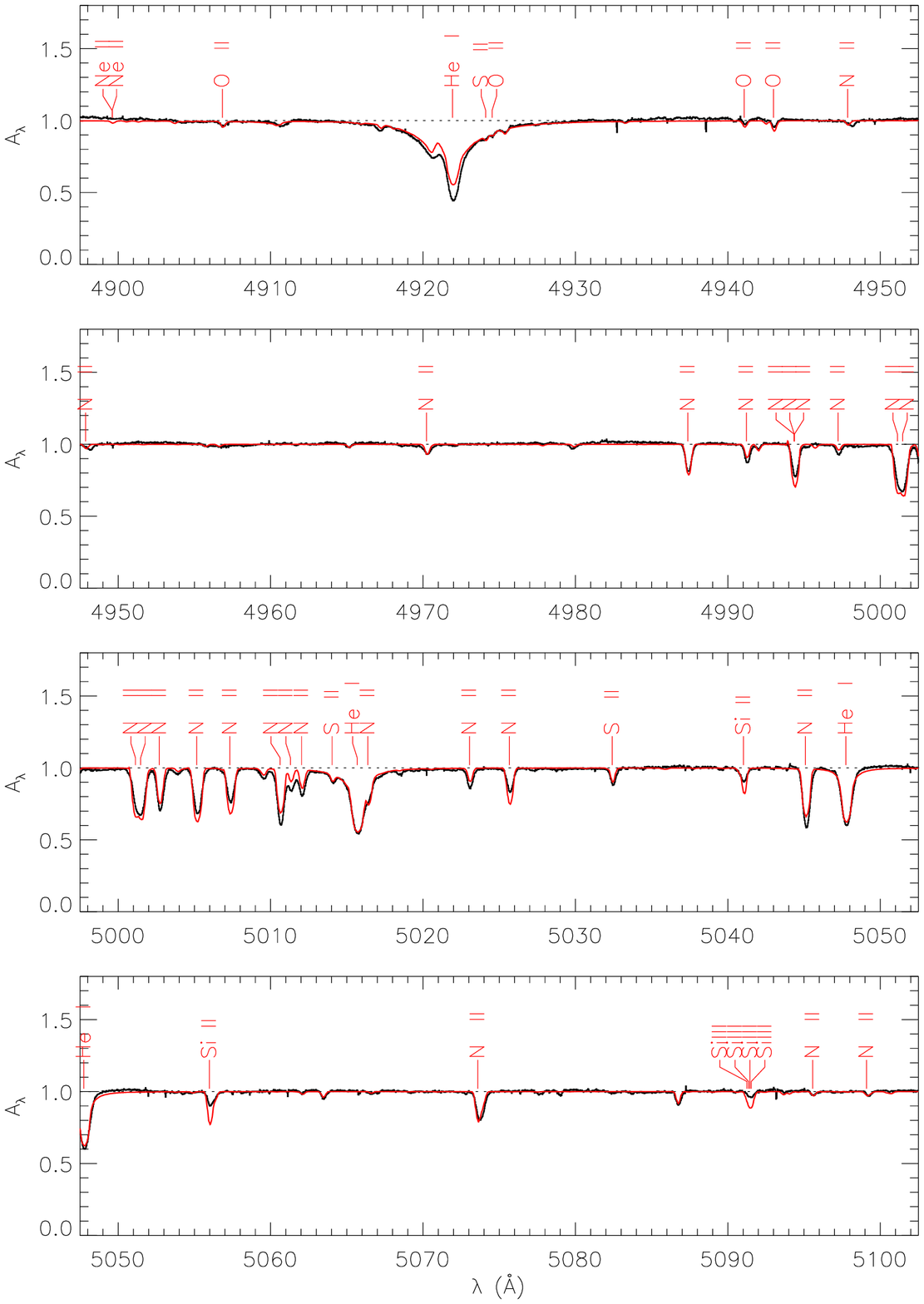}
\caption{(e) contd. } \label{f:atlas_e}
\end{figure*}   
\begin{figure*}\addtocounter{figure}{-1}
	\centering
		\includegraphics[width=0.90\textwidth,angle=0]{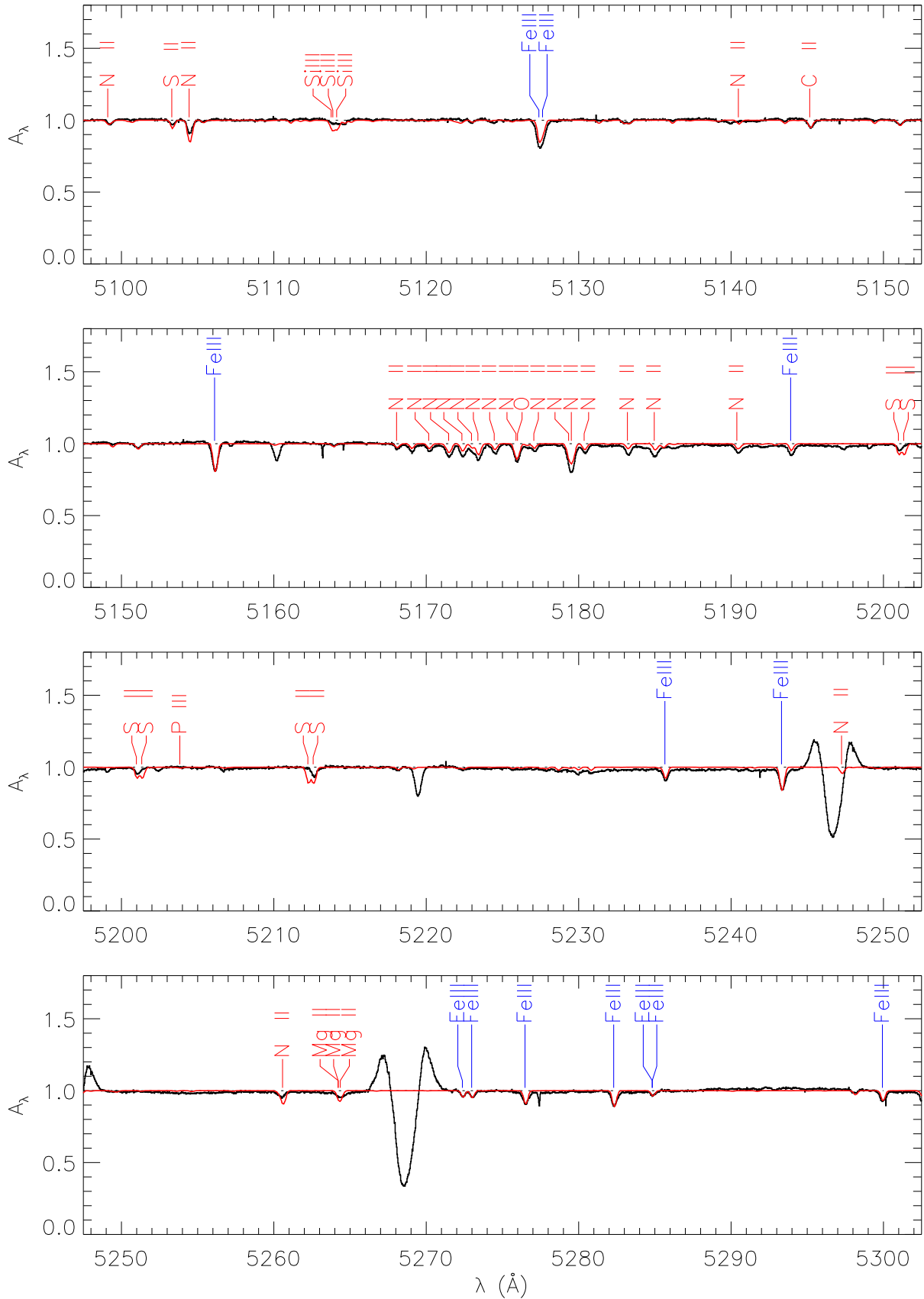}
\caption{(f) contd. Strong broad features around  5247 and 5268~\AA\  are instrumental artefacts.} \label{f:atlas_f}
\end{figure*}   
\begin{figure*}\addtocounter{figure}{-1}
	\centering
		\includegraphics[width=0.90\textwidth,angle=0]{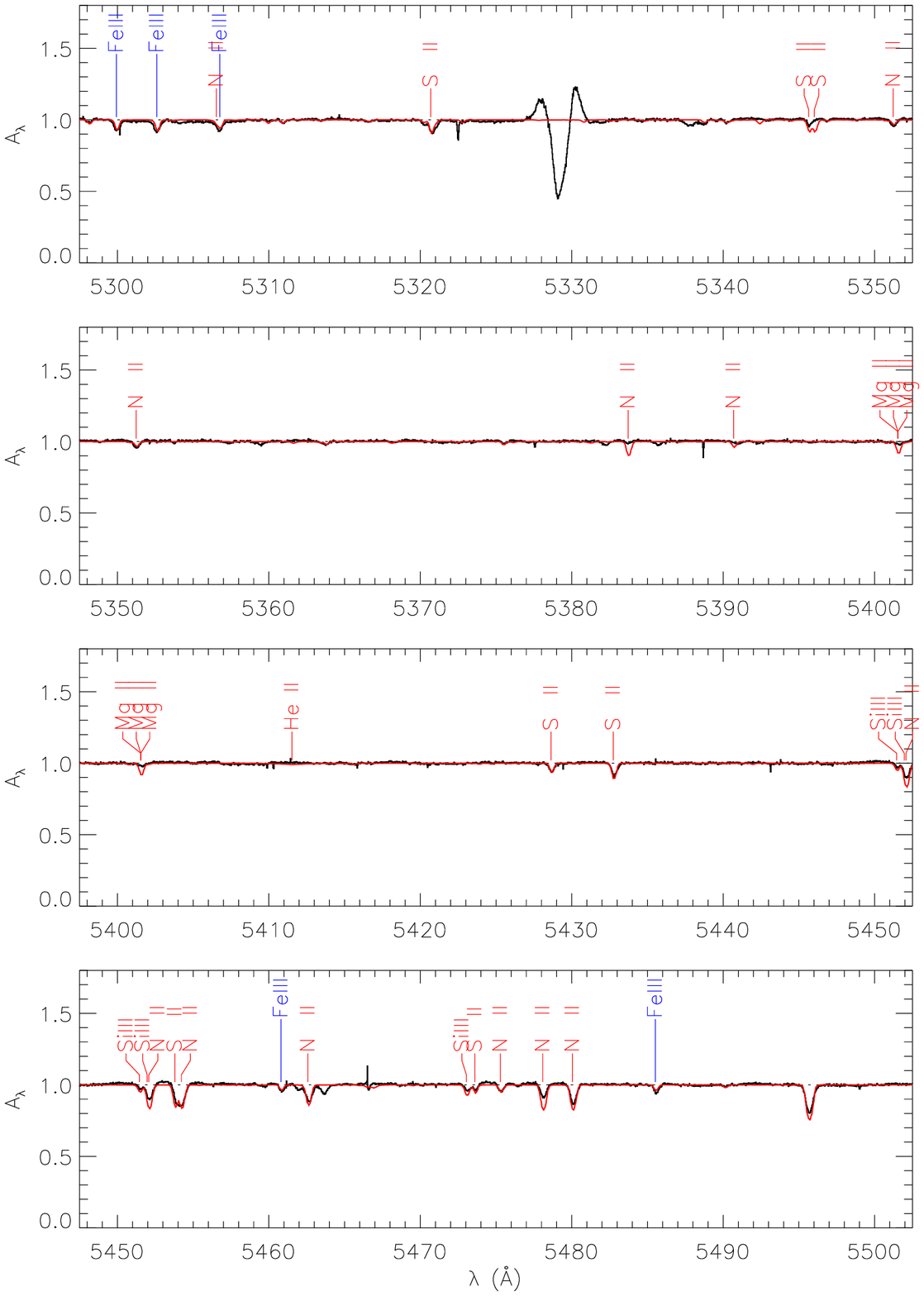}
\caption{(g) contd. A strong broad feature at 5329~\AA\ is an instrumental artefacts. } \label{f:atlas_g}
\end{figure*}   
\begin{figure*}\addtocounter{figure}{-1}
	\centering
		\includegraphics[width=0.90\textwidth,angle=0]{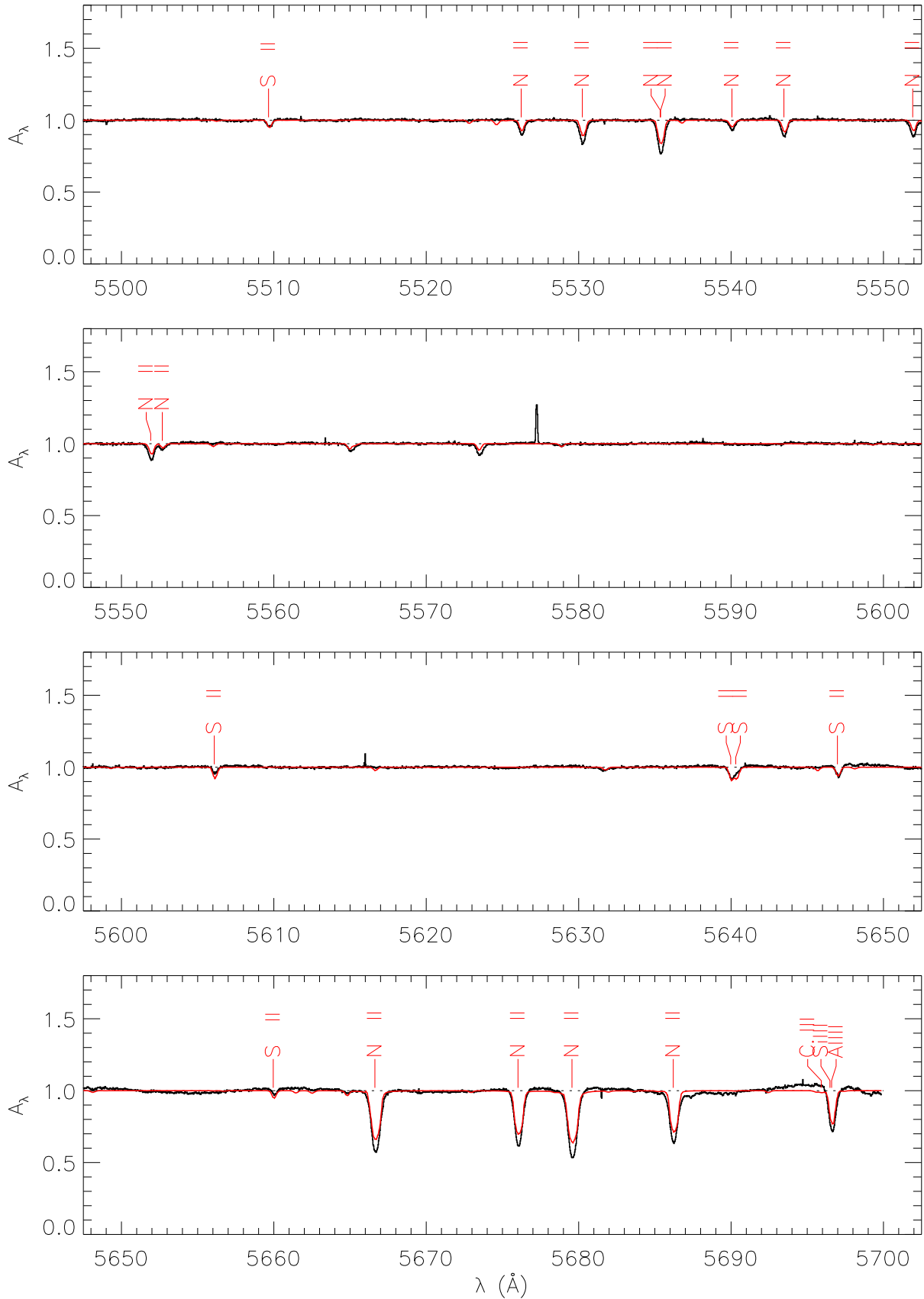}
\caption{(h) contd. An emission feature at 5577~\AA\ is a cosmic ray hit. } \label{f:atlas_h}
\end{figure*}

\label{lastpage}

\end{document}